\documentclass[12pt]{article}

\RequirePackage[OT1]{fontenc}
\RequirePackage{amsthm,amsmath,amssymb,amsfonts,graphicx,subfigure,bm,fullpage,setspace,color}
\usepackage{algpseudocode,algorithm}

\theoremstyle{plain}
                          \newtheorem{thm}{Theorem}

\theoremstyle{remark}         
\theoremstyle{definition} 
                          \newtheorem{exam}{Example}

\numberwithin{exam}{section}

\usepackage{mathtools}
\usepackage{rotating}

\newcommand\om{\Omega}
\newcommand\real{\mathbb{R}}

\newcommand\D{\mathcal{D}}

\newcommand\F{\mathcal{F}}

\newcommand\bX{\bm{X}}
\newcommand\bY{\bm{Y}}

\newcommand\bbet{\bm\beta}
\newcommand\bgam{\bm\gamma}
\newcommand\beps{\bm\epsilon}
\newcommand\blam{\bm\lambda}

\newcommand\M{\mathcal{M}}

\newcommand\BF{{\rm BF}}
\newcommand\corr{{\rm corr}}

\newcommand\bzero{\bm{0}}

\newcommand{\equalto}[2]{\underset{\textstyle\overset{\mkern4mu\parallel}{#2}}{#1}}

\def\I{{\mathbf{1}}}

\def\woMR#1{\w@MR#1MR#1MR\relax}%
\def\w@MR#1MR#2MR#3\relax{#2}

\def\@MR#1 #2\relax#3{%
 \href{http://www.ams.org/mathscinet-getitem?mr=#1}%
 {\MRfixed{#3}}}%

\def\MRfixed{MR\woMR}%

 \usepackage{natbib}
\usepackage{varioref}		
\labelformat{section}{Section~#1}
\labelformat{figure}{Figure~#1}
\labelformat{table}{Table~#1}

\title{Scalable Bayesian model averaging through local information propagation}
\author{Li Ma}

\begin{document}
\maketitle

\doublespace

\begin{abstract}
We show that a probabilistic version of the classical forward-stepwise variable inclusion procedure can serve as a general data-augmentation scheme for model space distributions in (generalized) linear models. This latent variable representation takes the form of a Markov process, thereby allowing information propagation algorithms to be applied for sampling from model space posteriors. In particular, we propose a sequential Monte Carlo method for achieving effective unbiased Bayesian model averaging in high-dimensional problems, utilizing proposal distributions constructed using local information propagation. We illustrate our method---called LIPS for local information propagation based sampling---through real and simulated examples with dimensionality ranging from 15 to 1,000, and compare its performance in estimating posterior inclusion probabilities and in out-of-sample prediction to those of several other methods---namely, MCMC, BAS, iBMA, and LASSO. In addition, we show that the latent variable representation can also serve as a modeling tool for specifying model space priors that account for knowledge regarding model complexity and conditional inclusion relationships.
\end{abstract}

\section{Introduction}
\vspace{-0.6em}

We consider Bayesian model averaging (BMA) \citep{hoeting:1999} in Gaussian linear regression though the methodology developed in this work is directly applicable to generalized linear models. Suppose the data are $n$ observations with $p$ potential predictors and a response.
Let $\bY=(y_1,y_2,\ldots,y_n)$ denote the response vector and $\bX_{j}=(x_{1j},x_{2j},\ldots,x_{nj})$ the $j$th predictor vector for $j=1,2,\ldots,p$. We consider linear regression models of the form
\vspace{-1.7em}

\[
\M_{\gamma}: \quad \bY = \I_{n}\alpha + \bX_{\gamma}\bbet_{\gamma}+\beps
\]
\vspace{-2.7em}

\noindent where $\I_{n}$ stands for an $n$-vector of ``1''s; $\beps=(\epsilon_1,\epsilon_2,\ldots,\epsilon_{n})$ is a vector of i.i.d.\ Gaussian noise with mean 0 and variance $1/\varphi$; $\gamma=(\gamma_1,\gamma_2,\ldots,\gamma_{p})\in \{0,1\}^{p}\equiv \om_{\M}$ is the model (identifier) vector, that is, a vector of indicators whose $j$th element $\gamma_j=1$ if and only if the $j$th variable $\bX_{j}$ enters the model; $\bX_{\gamma}$ and $\bbet_{\gamma}$ represent the corresponding design matrix and coefficients. Because $\gamma$ and the model $\M_{\gamma}$ are in one-to-one correspondence, we will use them interchangeably and refer to $\gamma$ simply as a model.

Let $\pi(\gamma)$ be the prior probability assigned to any model $\gamma\in\om_{\M}$. Then by Bayes' theorem, the posterior model probability is given by
\vspace{-1.4em}

\[
\pi(\gamma|\D) = \frac{\pi(\gamma)p(\D|\gamma)}{\sum_{\gamma\in \om_{\M}}\pi(\gamma)p(\D|\gamma)}
\]
\vspace{-2.1em}

\noindent where $p(\D|\gamma)$ is the marginal likelihood under $\M_{\gamma}$. More specifically,
\vspace{-1.6em}

\[
p(\D|\gamma) = \int p(\D|\theta_{\gamma},\gamma)\pi(\theta_{\gamma}|\gamma)d\theta_{\gamma}
\]
\vspace{-2.6em}

\noindent where $\theta_{\gamma}=(\alpha,\bbet_{\gamma},\varphi)$, representing the parameters given model $\M_{\gamma}$, and $\pi(\theta_{\gamma}|\gamma)$ is the prior on $\theta_{\gamma}$ given $\M_{\gamma}$. The notation $\pi(\theta_{\gamma}|\gamma)$ indicates that the prior on the coefficients can depend on the model under consideration. The integral can be evaluated analytically for many common priors and can often be approximated by Laplace approximation otherwise \citep{liang:2008}. 

BMA then concerns the prediction or estimation of some quantity of interest $\Delta$, and suggests using the corresponding posterior mean
\vspace{-1.5em}

\[
E(\Delta\,|\,\D)=\sum_{\gamma\in \om_{\M}} E(\Delta\,|\,\gamma,\D) \pi(\gamma|\D),
\]
\vspace{-2em}

\noindent as the predicted value. Extensive studies on BMA show that it has various desirable theoretical properties in terms of predictive performance, including minimizing the mean squared error and being optimal under the log-score criterion \cite[Sec.~2]{raftery:2003}.

In low-dimensional problems with less than $30$ variables, it is possible to carry out BMA exactly by enumerating the model space and computing the model space posterior $p(\gamma|D)$ exhaustively. Modern applications often involve many more variables, however, for which model space enumeration is infeasible. In such cases BMA is carried out through Monte Carlo and its efficacy relies critically on effectively sampling from the posterior on the large model space.
Tremendous advance has been made in the development of sampling and search algorithms on large model spaces. 
Some notable examples include (i)~efficient Markov Chain Monte Carlo (MCMC) algorithms---for example those developed in \cite{madigan:1995,geweke:1996,smith:1996,liang:2000,nott:2004,nott:2005,bottolo:2010,wilson:2010,ghosh:2011}; (ii) stochastic search algorithms---for example \cite{george:1993,george:1997,jones:2005,berger:2005,hans:2007,jasra:2007,scott:2008,shi:2011}; (iii)~alternative sampling strategies---for example \cite{clyde:2011,schaffer:2013,rockova:2014}.

In problems with more than hundreds of variables, efficient exploration over the vast $2^p$ model space is particularly challenging. In such cases, MCMC-based methods often have difficulty in convergence and mixing. In recent years, several scalable stochastic search and adaptive sampling algorithms have been proposed to tackle this challenge---see for example \cite{berger:2005,hans:2007,scott:2008,clyde:2011}. Rather than sampling from the actual posterior, these methods aim at searching the parts of the model space deemed to contain the ``best'' models. The idea is quite natural---if but a tiny fraction of the models can be explored in practical time, one might as well focus on the models best supported by the data. However, because the models sampled by these methods do not arise from the actual posterior, sample averages are biased estimates for the posterior mean \citep{heaton:2010,clyde:2012}.

 When prediction or estimation is the primary inferential goal, unbiased estimation of posterior means according to the BMA recipe is often desirable. However, for stochastic search and adaptive sampling algorithms, the sampling distribution of the models is typically unknown, and therefore one cannot easily correct the sampling bias \citep{clyde:2012}.

In this work we introduce a data augmentation scheme and use it to design a new sampling method for achieving (approximately) unbiased BMA in high-dimensional problems. Our starting point, however, may seem implausibly simple: the familiar {\em forward-stepwise variable selection procedure} taught in every first course on regression.
We show that a probabilistic version of this procedure provides a latent variable representation for all model space distributions.
This representation takes the form of a Markov process, and thus it allows BMA to be carried out through information propagation (forward-summation-backward-sampling) algorithms \citep[Sec.~2.4]{liu:2001}. In high-dimensional problems where exact information propagation is infeasible, we construct a sequential Monte Carlo (SMC) sampler \citep{liu:2001,delmoral:2006,cappe:2007} using $k$-step local information propagation based proposals, which achieves very efficient scalable unbiased BMA.

As a desirable side-product, we show that the Markov latent representation can also be used as a modeling tool for specifying model space priors. In particular, we show that prior knowledge regarding model complexity and conditional inclusion relationships among the predictors can be more conveniently specified under this representation than directly on the original model space. We argue that proper specification on these aspects of the model space prior is important in high-dimensional settings to address multiple testing.

The rest of the work is organized as follows. In \ref{sec:model} we introduce the probabilistic forward-stepwise (pFS) representation as a data augmentation scheme for model space distributions. We show that all model space distributions can be augmented this way, and show how to use this representation as a tool for specifying model space priors and provide the reasons why it is useful for addressing multiplicity in high-dimensional settings. We establish the Markov nature of the pFS representation, and construct a forward-backward information propagation algorithm for finding pFS representation of model space posteriors. We then consider high-dimensional settings where exact information propagation is infeasible, and present an SMC sampler for BMA that operates in the augmented model space using local information propagation proposals.
In \ref{sec:examples} we apply our method to several real and simulated examples with dimensionality ranging from 15 to 1,000, and investigate its performance for estimation and prediction in comparison to Markov Chain Monte Carlo \citep{madigan:1995,brown:1998,fernandez:2001}, Bayesian adaptive sampling \citep{clyde:2011}, iterated BMA \citep{annest:2009}, and LASSO \citep{tibs:2011}. We conclude in \ref{sec:discussion} with some discussion.

We close this introduction by relating the current work to some particularly relevant literature. It is worth noting that both the idea of data augmentation and that of SMC have been considered (separately) for Bayesian model choice and averaging by other authors. In particular, \cite{ghosh:2011} proposed a data augmentation technique to transform a general design matrix into an orthogonal one, which allows posterior sampling to be achieved much more effectively.
\cite{schaffer:2013} proposed a general SMC algorithm for Bayesian variable selection that uses a geometric bridge \citep{gelman:1998,chopin:2002} to create a path of target distributions, and use the binary nature of each dimension in the model space, through the ``logistic conditionals'' model, to construct proposals. In comparison, the data augmentation technique we introduce substantially simplifies the technicality of the SMC sampler. Finally, \cite{annest:2009} also considered the problem of scalable BMA. Their method---the iterated BMA---iteratively screens the predictors down to a small number ($\sim 30$), and carries out exact BMA on this reduced model space.
\vspace{-1.5em}

\section{Markov augmentation for model space distributions}
\label{sec:model}
\vspace{-0.7em}

We start with some notation.
 For any $\gamma\in \om_{\M}$, let $\gamma^{+j}\in \om_{\M}$ for $j\in\{1,2,\ldots,p\}$ denote the model with an additional variable $\bX_j$ included. Mathematically, $\gamma^{+j}$ is such that
$\gamma^{+j}_{l}=\gamma_{l}$ for $l\in \{1,2,\ldots,p\}\backslash\{j\}$ and $\gamma^{+j}_{j}=1$.
If $\gamma$ already contains $\bX_j$, that is $\gamma_j=1$, then $\gamma^{+j}=\gamma$.
\vspace{-3em}

\subsection{The probabilistic forward-stepwise representation}
\label{sec:fs_dist}

In the classical FS variable selection procedure, a regression model is built sequentially by adding one variable at a time until certain stopping criterion is met. At each step of the procedure, the statistician makes two {\em decisions}: (i) a {\em stopping} decision---whether to terminate the procedure and (ii) a {\em selection} decision---if we do not stop, which variable to include next. This procedure is deterministic---with the same stopping and selection criteria, two statisticians working on the same data will end up with exactly the same result.

Now we introduce a {\em randomized} procedure---or a probabilistic model---for generating regression models that imitates the classical FS procedure. This {\em probabilistic forward-stepwise} (pFS) procedure is obtained by randomizing the stopping and selection decisions.

\vspace{0.5em}

{\em The probabilistic forward-stepwise procedure}. The procedure also starts with the null model and adds variables one-at-a-time until stopping. It consists of $p$ steps of operation. Let $\gamma_{(t)}=(\gamma_{(t),1},\gamma_{(t),2},\ldots,\gamma_{(t),p}) \in \om_{\M}$ denote the {\em tentative} model after the $t$th step. To begin, we have $t=0$ and $\gamma_{(0)}=(0,0,\ldots,0)$, the null model. At the end of the $p$ steps, we have the {\em final} model $\gamma_{(p)}$. Below is an illustration of the flow of the procedure
\vspace{-1.5em}

\[
\gamma_{(0)}\rightarrow \gamma_{(1)} \rightarrow \cdots \rightarrow \gamma_{(t)} \rightarrow \cdots \rightarrow \gamma_{(p)}.
\]
\vspace{-2.5em}

Each of the $p$ steps can be described inductively as follows. Suppose we have completed the first $(t-1)$ steps. Then in the $t$th step, if the procedure has not ``stopped''---what ``stopping'' means will be described shortly, then first we draw a Bernoulli random variable
\vspace{-1.5em}

\[ S_{t} \sim {\rm Bernoulli}\left(\rho(\gamma_{(t-1)})\right)\]
\vspace{-2.5em}

\noindent where, as the notation indicates, the ``success'' probability $\rho(\gamma_{(t-1)})$ can depend on $\gamma_{(t-1)}$. If $S_t=1$, then the procedure {\em stops} in the $t$th step and we set $\gamma_{(t)}=\gamma_{(t-1)}$. Thus we call $\rho(\gamma_{(t-1)})$ the {\em stopping probability}. If instead $S_{t+1}=0$, then the procedure is not stopped in the $t$th step. In this case, we {\em randomly} select a new variable to add into the model. More specifically, we draw a multinomial Bernoulli variable $J_{t}$ from $\{1,2,\ldots,p\}$
\vspace{-1.5em}

\[
J_{t} \sim \text{Multi-Bern}\left(\{1,2,\ldots,p\};\blam(\gamma_{(t-1)})\right),
\]
\vspace{-2.5em}

\noindent where $\blam(\gamma_{(t-1)})=\left(\lambda_1(\gamma_{(t-1)}),\lambda_2(\gamma_{(t-1)}),\ldots, \lambda_p(\gamma_{(t-1)})\right)$ represent the {\em selection probabilities} with $\lambda_j(\gamma_{(t-1)})$ for each $j=1,2,\ldots,p$ being the probability for $J_t=j$. The selection probabilities also can depend on the model at the previous step $\gamma_{(t-1)}$. If $J_t=j$, we add the $j$th predictor into the model, and get $\gamma_{(t)}=\gamma^{+j}_{(t-1)}$.

On the other hand, if the procedure has already stopped in the first $t-1$ steps,
then in the $t$th step the procedure remains stopped---we set $S_{t}= 1$ and $\gamma_{(t)}=\gamma_{(t-1)}$. In this case, the selection variable $J_{t}$ does not matter at all, but for sake of completeness, we still assume that they are distributed as the multinomial Bernoulli given above. This completes our inductive description for the $t$th step of the procedure for $t=1,2,\ldots,p$.

\vspace{0.5em}

The final model $\gamma_{(p)}$ is determined by $S_1,J_1,S_2,J_2,\ldots,S_p,J_p$, which we call the (latent) {\em decision} variables. A pFS procedure is specified by their distributions through two mappings
\vspace{-3.5em}

\[
\rho: \om_{\M}\backslash\{(1,1,\ldots,1)\} \rightarrow [0,1] \quad \text{and} \quad \blam: \om_{\M}\backslash\{(1,1,\ldots,1)\} \rightarrow [0,1]^{p}
\]
\vspace{-2.5em}

\noindent where $\om_{\M}\backslash\{(1,1,\ldots,1)\}$ is the set of all non-full models. For each non-full $\gamma$, $\blam$ satisfies
\vspace{-2.5em}

\[
\lambda_j(\gamma)\geq 0 \text{ for all $j=1,2,\ldots,p$, } \quad \sum_{l=1}^{p}\lambda_l(\gamma)=1, \text{ and } \lambda_j(\gamma)=0 \text{ for all $j$ such that $\gamma_{j}=1$.}\]
\vspace{-2.5em}

\noindent These constraints ensure that the multinomial Bernoulli distribution for each $J_t$ is well-defined, and that only variables not in the tentative models will be selected.

The following theorem establishes the Markov property of the pFS procedure.
\begin{thm}[Markov property]
\label{thm:markov}
The sequence of tentative models $\{\gamma_{(t)}\}_{t=0}^{p}$ form a Markov chain,
and the transition probabilities are given by
\begin{align}
\pi_{t|t-1}\left(\gamma_{(t)}\, \big|\, \gamma_{(t-1)}\right)=\begin{cases}
1 & \text{if $|\gamma_{(t-1)}|<t-1$ and $\gamma_{(t)}=\gamma_{(t-1)}$,}\\
\rho\left(\gamma_{(t-1)}\right) & \text{if $|\gamma_{(t-1)}|=t-1$ and $\gamma_{(t)}=\gamma_{(t-1)}$,} \\
\left(1 - \rho\left(\gamma_{(t-1)}\right)\right)\cdot \lambda_{j}\left(\gamma_{(t-1)}\right) & \text{if $|\gamma_{(t-1)}|=t-1$ and $\gamma_{(t)}=\gamma^{+j}_{(t-1)}$,}\\
0 & \text{otherwise.}
\end{cases}
\end{align}
\vspace{-3.5em}

\noindent for $t=1,2,\ldots,p$ and $j=1,2,\ldots,p$.
\end{thm}
\vspace{-1em}

\begin{proof}[Proof]
See Online Supplementary Materials~S1.
\end{proof}

\begin{exam}
Let us consider the case with $p=5$. Suppose the decision variables take the following values: $S_1=0$, $J_1=4$, $S_2=0$, $J_2=2$, $S_3=0$, $J_3=5$, and $S_4=1$, then the sequence of tentative models are
\vspace{-3.5em}

\[ \equalto{(0,0,0,0,0)}{\gamma_{(0)}} \rightarrow \equalto{(0,0,0,1,0)}{\gamma_{(1)}} \rightarrow \equalto{(0,1,0,1,0)}{\gamma_{(2)}} \rightarrow \equalto{(0,1,0,1,1)}{\gamma_{(3)}} \rightarrow \equalto{(0,1,0,1,1)}{\gamma_{(4)}} \rightarrow \equalto{(0,1,0,1,1)}{\gamma_{(5)}}.\]
\vspace{-2em}

\noindent Note that because the procedure has stopped in Step~4, the values of $J_4$ and $J_5$ do not matter. The probability to get this sequence is
\vspace{-3.4em}

\begin{align*}
\left(1-\rho\left(0,0,0,0,0\right)\right)&\cdot \lambda_{4}\left(0,0,0,0,0\right) \times \left(1-\rho\left(0,0,0,1,0\right)\right) \cdot \lambda_2\left(0,0,0,1,0\right)\\
\times &\left(1-\rho\left(0,1,0,1,0\right)\right) \cdot \lambda_5\left(0,1,0,1,0\right)
\times \rho\left(0,1,0,1,1\right).
\end{align*}
\vspace{-3.4em}

\noindent We emphasize the difference between this probability and the {\em marginal} probability for the final model $\gamma_{(5)}$ to be $(0,1,0,1,1)$. To get the same final model, one can include the three variables $X_2$, $X_4$, and $X_5$ in any of the six possible orderings. For example, the probability of the chain that first includes $X_2$, then $X_5$, and finally $X_4$ before stopping is given by
\vspace{-3.4em}

\begin{align*}
\left(1-\rho\left(0,0,0,0,0\right)\right)&\cdot \lambda_{2}\left(0,0,0,0,0\right) \times \left(1-\rho\left(0,1,0,0,0\right)\right) \cdot \lambda_5\left(0,1,0,0,0\right)\\
\times &\left(1-\rho\left(0,1,0,0,1\right)\right) \cdot \lambda_4\left(0,1,0,0,1\right)
\times \rho\left(0,1,0,1,1\right),
\end{align*}
\vspace{-3.4em}

\noindent which may differ from the probability of the previous chain depending on $\rho$ and $\blam$. The marginal probability for $\gamma_{(5)}=(0,1,0,1,1)$ is the sum of the six probabilities of this form.
\end{exam}

Let $\pi$ be a probability measure on the model space $\om_{\M}$. If there exists a pFS procedure such that $\pi$ is the marginal distribution of the final model $\gamma_{(p)}$, then the pFS procedure is said to be a {\em pFS representation} for $\pi$.
Our next theorem shows that {\em all} model space distributions have pFS representations, and thus the pFS procedure can serve as a general data-augmentation tool, or latent variable representation, for model space distributions.

\begin{thm}[Generality]
\label{thm:generality}
All probability measures on $\om_{\M}$ have pFS representations.
\end{thm}
\begin{proof}
See Online Supplementary Materials~S1.
\end{proof}

This theorem implies that
{\em any} model space prior can be augmented by a pFS procedure with appropriately chosen $\rho$ and $\blam$ mappings. In the proof of Theorem~2, we provide a general recipe for finding a pFS representation for any model space prior. We will devote Section~\ref{sec:prior_spec} to show how to use the pFS representation as a tool for choosing model space priors that take into account prior information regarding model complexity and conditional inclusion relationships, both of which are important in high-dimensional problems for addressing multiplicity.

The Markov property of the pFS representation has an important implication---inference on the posterior model space distributions can be carried out through information propagation, or forward-summation-backward-sampling algorithms \citep[Sec.~2.4]{liu:2001}. Our next theorem 
shows how to find pFS representations for model space {\em posteriors} through information propagation.

\begin{thm}[Information propagation]
\label{thm:posterior_fs}
If a model space prior $\pi$ has a pFS representation with mappings $\rho$ and $\blam$, then the corresponding model space posterior given data $\D$ has a pFS representation with mappings $\rho(\cdot\,|\,\D)$ and $\blam(\cdot\,|\,\D)$ given as follows. For any $\gamma\in\om_{\M}\backslash\{1,1,\ldots,1\}$ and $j=1,2,\ldots,p$,
\begin{itemize}
\item if $\rho(\gamma)<1$,
\vspace{-3.5em}

\[
\rho(\gamma\,|\,\D) = \rho(\gamma)\cdot \BF_{0}(\gamma)/\phi(\gamma) \quad \text{and} \quad \lambda_j(\gamma\,|\,\D)=\lambda_j(\gamma)\cdot \frac{(1-\rho(\gamma))\cdot\phi(\gamma^{+j})}{\phi(\gamma)-\rho(\gamma)\cdot \BF_{0}(\gamma)}
\]
\vspace{-3.5em}

\item if $\rho(\gamma)=1$,
\vspace{-3em}

\[ \rho(\gamma\,|\,\D) = 1 \quad \text{and} \quad  \lambda_j(\gamma\,|\,\D)=\lambda_j(\gamma), \]
\end{itemize}
where $\BF_{0}(\gamma)=p(\D\,|\,\gamma)/p(\D\,|\,\bzero)$ is the Bayes factor (BF) of the model $\gamma$ with respect to the null model $\bzero=(0,0,\ldots,0)$,
and $\phi:\om_{\M}\rightarrow \real$ is a mapping defined recursively as follows
\vspace{-1em}

\[
\phi(\gamma)=\begin{cases} \BF_{0}(\gamma) & \text{if $|\gamma|=p$,} \\
\rho(\gamma)\cdot \BF_{0}(\gamma) + (1-\rho(\gamma))\cdot \sum_{j:\gamma_j=0}\lambda_j(\gamma)\phi(\gamma^{+j}) & \text{if $|\gamma|<p$.}
\end{cases}
\]
\end{thm}
\noindent Remark~I: Note that
the mapping $\phi$ is {\em recursive} in the sense that for each non-full model $\gamma$, $\phi(\gamma)$ is computed based on models including one more predictor. So once we have computed $\phi(\gamma)$ for all $\gamma$ of size $t\in\{1,2,\ldots,p\}$, we can compute $\phi(\gamma)$ for all $\gamma$ of size $t-1$.
\vspace{0.3em}

\noindent Remark~II: The Bayes factor, BF$_{0}(\gamma)$,
can be computed either exactly or through Laplace approximation for many common priors on regression parameters. We give explicit formulas for two such priors---Zellner's $g$ \citep{zellner:1981} and the hyper-$g$ \citep{liang:2008} in Online Supplementary Materials~S2.
\vspace{0.3em}

\noindent Remark~III: The pFS representation and the above theorem work for generalized linear models as well. The only complication is that one will again need numeric methods such as Laplace approximation to compute the Bayes factors.

\begin{proof}[Proof of Theorem~\ref{thm:posterior_fs}]
See Online Supplementary Materials~S1.
\end{proof}
\vspace{-0.5em}

The computation of $\phi$ in the above theorem corresponds to the ``forward summation'' step of information propagation.
On the other hand, the updating formulas for the posterior parameters correspond to the ``backward sampling'' stage of the information propagation.

The careful reader may notice that Theorem~\ref{thm:posterior_fs} by itself does not address the dimensionality problem because the recursion for $\phi$ requires an enumeration of the model space.
However, as we will see in Section~\ref{sec:smc} that a local version of the information propagation algorithm can help us construct highly effective proposal distributions for sequential importance sampling of high-dimensional model space posteriors.

\vspace{-1em}

\subsection{Specifying model space priors through pFS representation}
\label{sec:prior_spec}
\vspace{-0.5em}

While the main point of this work is to use the pFS representation to design a scalable method for carrying out BMA, we take a slight diversion in this section and show that the pFS representation can serve as a useful modeling tool for specifying model space priors as well.
This is because the two mappings---$\rho$ and $\blam$---correspond to two essential aspects of model space priors. More specifically, $\rho$ characterizes prior information on the model complexity, whereas $\blam$ the relative importance of the predictors.
We next elaborate on each.

The mapping $\rho$ encodes the prior information about model complexity. A convenient but flexible way to specify $\rho$ is to let $\rho(\gamma)$ depend on $\gamma$ through the model size $|\gamma|$, that is,
\vspace{-1.5em}

\[ \rho(\gamma)=h(|\gamma|) \quad \text{where $h:\{0,1,\ldots,p-1\}\rightarrow [0,1]$.} \]
\vspace{-2.5em}

\noindent Intuitively, $h(s)$ is the prior probability that the model is of size $s$ given that its size $\geq s$.

Any prior marginal distribution on the model size can be specified through proper choice of the function $h$. More specifically, suppose one wishes to specify a prior distribution $(q_0,q_1,q_2,\ldots,q_p)$ on the model size where $q_{s}$ is the prior probability for $|\gamma|=s$ for $s=0,1,2,\ldots,p$. Then one can let
\[
h(s) = \frac{q_{s}}{1-\sum_{r=0}^{s-1} q_{r}}.
\]
In particular, for $q_0=q_1=\cdots = q_p=1/(p+1)$, we can let $h(s)=1/(p+1-s)$. \cite{scott:2010} show that this choice helps achieve effective control on ``false inclusions''.

The prior selection probability mapping $\blam$ encodes the prior information regarding the relative importance of the predictors. If no such prior information is available, a simple choice is the uniform selection probabilities, that is, to let
\vspace{-1.7em}

\[ \lambda_j(\gamma)\equiv 1/(p-|\gamma|)\quad \text{for all $j$ such that $\gamma_j=0$.}\]
\vspace{-2.6em}

\noindent In many applications, one does have prior knowledge regarding the relative importance of the predictors. For example, in genetic applications where the predictors are gene markers, the biologist may know that some markers are more likely to be related to the response (e.g.\ a phenotype measurement) than others. For instance, based on past experience one may believe that the markers lying in genes are more likely to be associated with the response than those outside of genes. One can then choose $\blam$ such that
\vspace{-1.2em}

\[
\lambda_j(\gamma) = \begin{cases} c\cdot k(\gamma)/(p-|\gamma|) & \text{if $\gamma_j=0$ and marker~$j$ is in a gene,}\\
k(\gamma)/(p-|\gamma|) & \text{if $\gamma_j=0$ and marker~$j$ is not in a gene,}\\
0  & \text{if $\gamma_j=1$}\end{cases}
\]
\vspace{-1.5em}

\noindent where $c\geq 1$ is a constant that specifies how much more likely are markers in genes to be included, and $k(\gamma)$ is a constant that ensures $\sum_{j}\lambda_j(\gamma)=1$. In this particular example, $k(\gamma)$ depends on the number of markers in (and out) of genes that are not already in $\gamma$.

Sometimes one has prior knowledge regarding the importance of predictors {\em in relation to} the other variables in the model. A predictor $X_i$ may be more (or less) likely to be in the model when another set of variables are also included. For example, gene markers belonging to the same biological pathway may be expected to enter the model together. Incorporating such conditional knowledge is often difficult when specifying a model space prior directly, but is made straightforward under the pFS representation. If say, markers $i$, $j$, and $k$ are in the same pathway then we can let
\vspace{-1.2em}

\[
\lambda_j(\gamma) = \begin{cases} c\cdot k(\gamma)/(p-|\gamma|) & \text{if $\gamma_j=0$, and $\gamma_i=1$ or $\gamma_k=1$,}\\
k(\gamma)/(p-|\gamma|) & \text{if $\gamma_j=0$, and $\gamma_i=\gamma_k=0$,}\\
0 & \text{if $\gamma_j=1$} \end{cases}
\]
\vspace{-1.5em}

\noindent for some $c>1$, and again $k(\gamma)$ is a known constant that ensures $\sum_{j}\lambda_j(\gamma)=1$. When some of the predictors represent interactions, a common constraint is to allow potential inclusion of the interaction terms only when all of the corresponding main effects are already in the model. This can be achieved easily under the pFS representation. For example, if $X_k$ is the interaction term between $X_i$ and $X_j$, then this constraint can be imposed by setting $\lambda_k(\gamma)=0$ if $\gamma_i=0$ or $\gamma_j=0$.

In some applications it is desirable to include variables in blocks. For example, if some genes work together, then one can enforce this block of variables to be included or excluded together. One can achieve such whole block inclusion under the current one-variable-at-a-step formulation by specifying $\rho$ and $\blam$ such that if one of the variables in a block is included, then $\rho(\gamma)=0$ and $\blam(\gamma)$ puts all mass on the other variables in the block until all variables in the block are included. Blocked inclusion is particularly useful when the individual effects of some predictors are very small. In this case, it helps combine information across multiple predictors to increase the statistical power for detecting such effects.

A general challenge in high-dimensional inference is proper adjustment for multiplicity (or multiple testing). In the context of Bayesian model choice and averaging, multiplicity is reflected in the spread of the model space prior over a tremendous number of models, many of which are ``bad''.  In addressing multiplicity, it is important to control the rate of ``false'' inclusions of irrelevant predictors while maximizing the power or sensitivity for recovering ``true'' inclusions of relevant ones. It has been shown that simple strategies such as assuming {\em a priori} each predictor is included independently with equal probability does not properly control for false inclusion \citep{scott:2010}. Stronger control on the model complexity is necessary. At the same time, to improve inference, one should incorporate as much relevant background knowledge as possible, thereby concentrating more prior probability mass on models that are more likely to be true. In this regard, the pFS representation allows specifying prior model complexity and relative importance of predictors in a decoupled, flexible fashion, and so is particularly desirable in high-dimensional settings. Online Supplmentary Materials~S3 gives another example that illustrates the flexibility in prior specification under the pFS representation---a strategy for addressing model space redundancy, the case when many predictors are highly collinear and redundant in the sense that including any one of them will capture virtually all of the association with the response \citep{george:1999,george:2010}.

{\em Specifying symmetric model space priors.} We have seen that the pFS representation provides an intuitive framework for specifying model space priors. Of course things can go the other direction too---one can start from a given model-space prior chosen by some other means and find a corresponding pFS representation. In the proof of Theorem~\ref{thm:generality}, we provide a general though complex recipe for finding a pFS representation for any model space distribution. For ``symmetric'' priors---those that assign equal probability to models of equal size, finding a pFS representation is extremely easy.
Specifically, for any symmetric prior $\pi$ we can let $\rho(\gamma)=h(|\gamma|)$ with $h(0)=\pi(0,0,\ldots,0)$, and inductively set
\[
h(t) = \frac{\pi(\gamma)\cdot {{p}\choose{t}}}{\prod_{s=1}^{t-1}\left(1-h(s)\right)} \quad \text{for $t=1,2,\ldots,p-1$}
\]
where $\gamma$ is any model of size $t$. In addition, we let $\lambda_j(\gamma)=1/(p-|\gamma|)\cdot \I(\gamma_j=0)$. One can verify that the marginal distribution of the final model is exactly~$\pi$.
\vspace{-1em}

\subsection{Scalable BMA through LIPS}
\label{sec:smc}
\vspace{-0.5em}

Next we shift from a modeling perspective to a computational focus. In particular, we use the pFS representation to construct a scalable algorithm for BMA. Our basic strategy is sequential importance sampling---drawing models from some sequentially constructed proposal distribution and using the (sequentially updated) ratio between the proposal and the  actual model space posterior
as weights to correct for the sampling bias.
The effectiveness of the sampling depends critically on the choice of the proposal. In high-dimensional settings,
choosing proposal distributions that are close to the model space posterior is especially important. To this end, the pFS representation provides a powerful and natural means to constructing effective proposals using local information propagation.

We now describe our main algorithm, called {\em local information propagation based sampling}, or LIPS, for drawing weighted samples from the model space posterior and constructing (approximately) unbiased BMA estimates.
We first describe the general schema of LIPS, and then provide details for each of its main components---particle propagation, weight updates,
and the construction of (approximately) unbiased estimates.
 Readers may directly refer to the box ``Algorithm~\ref{alg:pf}'' for the pseudo-code of the entire algorithm.

First we introduce some more notation. Let $\pi(\cdot)$ be our model space prior which has a pFS representation with mappings $\rho$ and $\blam$, and $\pi(\cdot\,|\,\D)$  the corresponding model space posterior. Next, for $t=0,1,2,\ldots,p$, we define $\pi_{0:t}(\cdot)$ as the model space prior induced by the pFS representation with mappings $\rho_{0:t}$ and $\blam_{0:t}$ defined as follows
\vspace{-1.2em}

\[
\rho_{0:t}(\gamma) = \begin{cases} \rho(\gamma) & \text{if $|\gamma| \leq t$ }\\
1 & \text{if $|\gamma| > t$}  \end{cases} \quad \text{and} \quad \blam_{0:t}(\gamma)=\blam(\gamma)
\]
\vspace{-1.2em}

\noindent for all $\gamma\in \om_{\M}$. In other words, $\pi_{0:t}(\cdot)$ is the corresponding model space prior that one would get if we force the pFS procedure for $\pi(\cdot)$ to stop at the $(t+1)$st step. The prior $\pi_{0:t}(\cdot)$ places all probability mass on models up to size $t$, and assigns probability in the same way as $\pi(\cdot)$ to models of size up to $t-1$, while concentrating the mass that $\pi(\cdot)$ places on models of size $\geq t$ on those of size $t$. In particular, we have $\pi_{0:p}(\cdot)=\pi(\cdot)$.

Accordingly, let $\pi_{0:t}(\cdot\,|\,\D)$ denote the model space posterior corresponding to the prior $\pi_{0:t}(\cdot)$ for $t=0,1,2,\ldots,p$.
The sequence of posteriors, $\{\pi_{0:t}(\cdot\,|\,\D):t=0,1,2,\ldots,p\}$, form a sequence of {\em target distributions} that approximate $\pi(\cdot\,|\,\D)$ in increasing degrees. More specifically, among models of size up to $t-1$, $\pi_{0:t}(\cdot\,|\,\D)$ assigns relative probability in the same way as $\pi(\cdot\,|\,\D)$. That is for two models $\gamma_1$ and $\gamma_2$ of sizes $\leq t-1$, $\pi_{0:t}(\gamma_1\,|\,\D)/\pi_{0:t}(\gamma_2\,|\,\D)= \pi(\gamma_1\,|\,\D)/\pi(\gamma_2\,|\,\D)$. In addition, the last target distribution $\pi_{0:p}(\cdot\,|\,\D)=\pi(\cdot\,|\,\D)$.

The general schema of LIPS is as follows. We {\em propagate} each of $N$ particles by simulating a pFS procedure. Thus after $t$ steps of propagation, the value of each particle is a tentative model of size up to $t$, and we {\em update} the importance weight (up to a normalizing constant) for the particles according to the current target $\pi_{0:t}(\cdot\,|\,\D)$.
After $p$ steps, we end up with a collection of final models, along with their importance weights corresponding to the last target distribution $\pi_{0:p}(\cdot\,|\,\D)=\pi(\cdot\,|\,\D)$. We use these weighted samples to carry out BMA by computing weighted averages in the form of Horvitz-Thompson (H-T) estimators \citep{horvitz:1952,clyde:2011}.

Next we elaborate on each component of LIPS.
Due to constraint of space, we do not provide all the relevant background on SMC. Interested readers may refer to \cite{liu:2001} for an excellent coverage on the topic.
\vspace{0.5em}

{\em Particle propagation}. We propagate $N$ particles in parallel by simulating $N$  pFS procedures. More specifically, for the $i$th particle where $i=1,2,\ldots,N$, we simulate a  sequence of tentative models
\vspace{-3em}

\[ \gamma^{i}_{(0)}\rightarrow \gamma^{i}_{(1)} \rightarrow \ldots \rightarrow \gamma^{i}_{(p)}\]
\vspace{-2.5em}

\noindent from a pFS procedure with mappings $\hat{\rho}(\cdot\,|\,\D)$ and $\hat{\blam}(\cdot\,|\,\D)$.
We call $\hat{\rho}(\cdot\,|\,\D)$ and $\hat{\blam}(\cdot\,|\,\D)$ the {\em proposal} mappings.  The ``hat'' notation reflect the fact that they typically are not exactly mappings for the actual posterior $\pi(\cdot\,|\,\D)$, but they can be chosen to approximate the latter. The ``$|\D$'' notation indicates that the proposal mappings may depend on the data.

From Theorem~\ref{thm:markov}, we know that the sequence  $\gamma^{i}_{(0)}, \gamma^{i}_{(1)},\ldots,\gamma^{i}_{(p)}$ form a Markov chain whose transition probability in the $t$th step is given by
\vspace{-3em}

\begin{align*}
q_{t|t-1}\left(\gamma^{i}_{(t)}\, \big|\, \gamma^{i}_{(t-1)},\D\right)=\begin{cases}
1 & \text{if $|\gamma^{i}_{(t-1)}|<t-1$ and $\gamma^{i}_{(t)}=\gamma^{i}_{(t-1)}$,}\\
\hat{\rho}\left(\gamma^{i}_{(t-1)}\,|\,\D\right) & \text{if $|\gamma^{i}_{(t-1)}|=t-1$ and $\gamma^{i}_{(t)}=\gamma^{i}_{(t-1)}$,} \\
\left(1 - \hat{\rho}\left(\gamma^{i}_{(t-1)}\,|\,\D\right)\right)\cdot \hat{\lambda}_{j}\left(\gamma^{i}_{(t-1)}\,|\,\D\right) & \text{if $|\gamma^{i}_{(t-1)}|=t-1$ and $\gamma^{i}_{(t)}=\gamma^{i,+j}_{(t-1)}$,}\\
0 & \text{otherwise.}
\end{cases}
\end{align*}
\vspace{-1.7em}

\noindent The mapping $q_{t|t-1}(\cdot\,|\,\cdot,\D): \om_{\M} \times \om_{\M}\backslash\{1,1,\ldots,1\} \rightarrow [0,1]$ is called the {\em proposal (transition) kernel}. The choice of the proposal kernel, or equivalently that of $\hat{\rho}$ and $\hat{\blam}$, is critical to ensuring the efficiency of the algorithm.

{\em Local information propagation.} The pFS representation
allows us to use {\em local} information propagation (LIP) to find effective proposals.
To see how to carry this out in practice, first for each model vector $\zeta \in \om_{\M}$, we define a mapping $\phi_{\zeta}:\om_{\M}\rightarrow \real$ such that for any $\gamma\in\om_{\M}$
\vspace{-5em}

\[
\phi_{\zeta}(\gamma) = \begin{cases} \BF_0(\gamma) & \text{if $|\gamma| \geq |\zeta|+k$ or $|\gamma|=p$,}\\
\rho(\gamma)\BF_0(\gamma) + (1-\rho(\gamma)) \cdot \sum_{j:\gamma_j=0}\lambda_j(\gamma)\,\phi_{\zeta}(\gamma^{+j}) & \text{otherwise.}
\end{cases}
\]
\vspace{-2em}

\noindent Next, we define a mapping $\hat{\phi}:\om_{\M}\rightarrow \real$ by setting $\hat{\phi}(\gamma)=\phi_{\gamma}(\gamma)$, that is by letting $\zeta=\gamma$, for all $\gamma\in \om_{\M}$. The mapping $\hat{\phi}$ approximates the mapping $\phi$ given in Theorem~\ref{thm:posterior_fs} by carrying out the recursion up to $k$ levels at a time.
Accordingly, we define the
proposal mappings $\hat{\rho}(\cdot\,|\,\D)$ and $\hat{\blam}(\cdot\,|\,\D)$ as follows. For any $\gamma \in \om_{\M}$ and $j=1,2,\ldots,p$,
if $\rho(\gamma)<1$,
\vspace{-2.5em}

\[
\hat{\rho}(\gamma\,|\,\D) =  \rho(\gamma)\cdot \BF_0(\gamma)/\hat{\phi}(\gamma) \quad \text{and} \quad \hat{\lambda}_j(\gamma\,|\,\D) =  \lambda_j(\gamma)\cdot \frac{\left(1-\rho(\gamma)\right)\cdot \hat{\phi}(\gamma^{+j})}{\hat{\phi}(\gamma)-\rho(\gamma)\cdot \BF_0(\gamma)},
\]
\vspace{-2.5em}

\noindent while if $\rho(\gamma)=1$, $\hat{\rho}(\gamma\,|\,\D) = 1$ and $\hat{\lambda}_j(\gamma\,|\,\D)=\lambda_j(\gamma)$. The larger $k$ is, the more closely the proposal mappings approximate the actual posterior mappings. 
In particular, if $k=p$ then the proposal mappings exactly recover the latter, though this is unachievable in high-dimensional problems.  We note that the above LIP proposal can be shown to be a special case of the so-called $k$-step look-ahead proposal \citep{lin:2013}.
\vspace{0.5em}

{\em Weight update}.  Next we derive the appropriate sequential weight updates for the particles.
To begin, set $w^{i}_{(0)}=1$ for all $i$.  Then for $t\geq 1$, by Bayes'~theorem
\vspace{-3em}

\begin{align}
\label{eq:target}
\pi_{0:t}\left(\gamma^{i}_{(t)}\,\big|\,\D\right)&\propto \pi_{0:t}\left(\gamma^{i}_{(t)}\right)\cdot p\left(\D\,\big|\,{\gamma^{i}_{(t)}}\right) \nonumber \\
&=\pi_{0:t-1}\left(\gamma^{i}_{(t-1)}\right)\cdot p\left(\D\,\big|\,{\gamma^{i}_{(t-1)}}\right)\cdot \pi_{t|t-1}\left(\gamma^{i}_{(t)}\big|\gamma^{i}_{(t-1)}\right)\cdot\frac{p\left(\D\,\big|\,{\gamma^{i}_{(t)}}\right)}{p\left(\D\,\big|\,{\gamma^{i}_{(t-1)}}\right)}  \nonumber \\
&=\pi_{0:t-1}\left(\gamma^{i}_{(t-1)}\right)\cdot p\left(\D\,\big|\,{\gamma^{i}_{(t-1)}}\right)\cdot \pi_{t|t-1}\left(\gamma^{i}_{(t)}\big|\gamma^{i}_{(t-1)}\right)\cdot \BF\left({\gamma^{i}_{(t)}},{\gamma^{i}_{(t-1)}}\right)  \nonumber \\
&\propto \pi_{0:t-1}\left(\gamma^{i}_{(t-1)}\,\big|\,\D\right)\cdot \pi_{t|t-1}\left(\gamma^{i}_{(t)}\big|\gamma^{i}_{(t-1)}\right)\cdot \BF\left({\gamma^{i}_{(t)}},{\gamma^{i}_{(t-1)}}\right)
\end{align}
\vspace{-2.5em}

\noindent where $\BF({\gamma^{i}_{(t)}},{\gamma^{i}_{(t-1)}}):=p(\D\,|\,{\gamma^{i}_{(t)}})/p(\D\,|\,{\gamma^{i}_{(t-1)}})$, and
$\pi_{t|t-1}(\gamma_{(t)}^{i}\,|\,\gamma_{(t-1)}^{i}):=\pi_{0:t}(\gamma^{i}_{(t)})/\pi_{0:t-1}(\gamma^{i}_{(t-1)})$, which is exactly the {\em prior} transition probability under the corresponding pFS representation for the prior as given in Theorem~\ref{thm:markov}.
Thus, by Eq.~\eqref{eq:target}, the weight for the $i$th particle after the $t$th step is updated to
\vspace{-3em}

\begin{align*}
w^{i}_{(t)}
=w^{i}_{(t-1)}\cdot \frac{\pi_{t|t-1}\left(\gamma^{i}_{(t)}\, \big|\, \gamma^{i}_{(t-1)}\right)}{q_{t|t-1}\left(\gamma^{i}_{(t)} \big| \gamma^{i}_{(t-1)}, \D \right)} \cdot \BF\left({\gamma^{i}_{(t)}},{\gamma^{i}_{(t-1)}}\right).
\end{align*}
\vspace{-2em}

\noindent Therefore, once we compute the BF between $\gamma^{i}_{(t)}$ and $\gamma^{i}_{(t-1)}$, we can update the weights sequentially. We provide the specific form of this BF under two common priors on the regression coefficients---the $g$-prior and the hyper-$g$ prior---in Online Supplementary Materials~S2.
\vspace{0.5em}

{\em BMA through H-T estimation.} At the end of each particle simulation, we end up with a final model $\gamma^{i}_{(p)}$ and a weight $w^{i}_{(p)}$. Suppose now we are interested in estimating the posterior mean of a quantity $\Delta$, that is $E(\Delta\,|\,\D)$, and conditional on any given model $\gamma$, $E(\Delta\,|\,\gamma,\D)$ can be evaluated. Then a consistent and approximately unbiased estimator, called the Horvitz-Thompson (H-T) estimator \citep{horvitz:1952,clyde:2012}, for $E(\Delta\,|\,\D)$ is given by
\[
\delta_{HT} = \frac{\sum_{i=1}^{N} w^{i}_{(p)} \cdot E( \Delta\,|\,\gamma^{i}_{(p)},\D)}{\sum_{i=1}^{N} w^{i}_{(p)}}.
\]

Let $W_i=w^{i}_{(p)}$, $\bar{W}=\sum_i W_i/N$, $Z_i =  w^{i}_{(p)}\cdot E(\Delta | \gamma^{i}_{(p)},\D)$, and $\bar{Z} = \sum_i Z_i / N$. Then we can estimate the variance of the $\delta_{HT}$ by \cite[p.35]{liu:2001}
\[
\widehat{{\rm Var}}(\delta_{HT}) = \frac{1}{N}\left[ \delta_{HT}^2\cdot \frac{\sum_{i=1}^{N} (W_i-\bar{W})^2}{N-1} +\frac{\sum_{i=1}^{N}(Z_i-\bar{Z})^2}{N-1} - 2 \delta_{HT} \cdot\frac{\sum_{i=1}^{N}(W_i-\bar{W})(Z_i-\bar{Z})}{N-1}\right].
\]
When multiple CPU cores are available or when the available computing memory is not enough for storing all $N$ particles, one can divide the $N$ total number of particles into $L$ ``islands'', each containing $N/L$ particles. One can then compute the H-T estimate using the particles in each island. We call the average of the $L$ estimators a {\em islanded (H-T) estimator}. Specifically, let $\delta_{HT,l}$ be the H-T estimator for the $l$th island, then the islanded estimator is given by $\bar{\delta}_{HT}=\sum_{l=1}^{L}\delta_{HT,l}/L$ and its variance can be estimated by $\sum_{l} (\delta_{HT,l}-\bar{\delta}_{HT})^2/(L(L-1))$.
The $L$ islands can be generated either on $L$ CPU cores in an (embarrassingly) parallel fashion, or sequentially on a single computing core using $N/L$ fraction of the required memory for storing $N$ particles.
This ``islanding'' idea for variance reduction and parallelization has been explored previously in the literature. See for example \cite{lak:2013}.

This completes the description of all the components in LIPS. Next we provide some general guidelines on the choice of the number of particles $N$ and the number of steps $k$.

{\em Choice of the number of particles.} There are two main considerations in choosing the number of particles $N$. The first is computational feasibility, which is determined by the amount of available RAM and computing time. In practice, one can take an initial run of the algorithm with a relatively small number of particles, and record the time and RAM used in the run. CPU time and RAM consumption are linear in the number of particles, and therefore one can compute the maximum number of particles allowed by the resources. The other consideration in choosing $N$ is the Monte Carlo accuracy attained in the estimates---if one has a desired level of precision in terms of the standard error then one can choose $N$ to achieve that accuracy. One can again use initial runs to attain rough estimates of the standard error of the H-T estimator, and then increase $N$ to the desired level. When achieving the lowest Monte Carlo error is desired, then one should adopt the maximum number of particles allowed. In applications where short computing time is of prime concern, one can choose the smallest number of particles necessary for achieving the desired level of precision, assuming it is achievable.

{\em Choice of $k$.} Generally, the larger $k$ is, the better the proposal approximates the posterior, and so fewer particles are needed to achieve the same precision, but with a larger $k$, more computation is incurred to propagate each particle. When $k=1$, the proposal is greedy and essentially based on the main effect of the predictors. Thus it is recommendable to choose $k\geq 2$ to take into account higher order effects in building the proposals. At the time of this writing, depending on the available computational resources, a general rule of thumb is that for problems with $p \leq 200\sim 500$, one can choose $k=3$, while for larger problems set $k=2$. 

\begin{algorithm}[p]
\caption{Local information propagation based sampling (LIPS)}
\label{alg:pf}

\begin{algorithmic}
\\
\For{$i=1,2,\ldots,N$}  \Comment{Initialization}
\State{Start from the null model: $\gamma^{i}_{(0)}=(0,0,\ldots,0)$.}
\State{Assign initial weights: $w^{i}_{0}=1$ and $\tilde{w}^{i}_{(0)}=1/N$.}
\EndFor
\\

\For{$t=1,2,\ldots,p$}\\

\For{$i=1,2,\ldots,N$}
\Comment{Propagation and weight update}
\State{Proposal construction:}
\[ \text{Find the proposal mappings $\hat{\rho}$ and $\hat{\blam}$ through LIP.} \]

\State{Particle Propagation:}
\vspace{-0.5em}

\[
\gamma^{i}_{(t)} \sim q_{t|t-1}\left(\cdot \,\big|\,\gamma^{i}_{(t-1)},\D\right).
\]
\State{Weight update:}
\vspace{-0.5em}

\[
w^{i}_{(t)}
=w^{i}_{(t-1)}\cdot \frac{\pi_{t|t-1}\left(\gamma^{i}_{(t)}\, \big|\, \gamma^{i}_{(t-1)}\right)}{q_{t|t-1}\left(\gamma^{i}_{(t)}\, \big|\, \gamma^{i}_{(t-1)}, \D \right)} \cdot \BF\left(\gamma^{i}_{(t)},\gamma^{i}_{(t-1)}\right).\]
\EndFor\\

\EndFor
\\
\Comment{Bayesian model averaging}
\State{Compute Horvitz-Thompson estimate for $E(\Delta\,|\,\D)$:}
\[
\delta_{HT} =  \frac{\sum_{i=1}^{N} w^{i}_{(p)} \cdot E( \Delta\,|\,\gamma^{i}_{(p)},\D)}{\sum_{i=1}^{N} w^{i}_{(p)}}.
\]
\end{algorithmic}
\end{algorithm}
\vspace{-1em}

\section{Numerical examples}
\label{sec:examples}
\vspace{-0.5em}

In this section we apply LIPS to several real and simulated data sets.Each data set consists of a response variable $\bY$ and a covariate design matrix $\bX$. We use LIPS to carry out two types of BMA estimation on linear models. The first is estimating the posterior inclusion probability (PIP) of the potential predictors, which is to set $\Delta = \I(\gamma_j=1)$ for $j=1,2,\ldots,p$. The other is for out-of sample prediction, which is to set $\Delta = Y_{new}$ where $Y_{new}$ is the response of a new observation with predictor values $\bX_{new}$. 

We start from a low-dimensional example (US crime data) in which the actual model space posterior can be computed through enumeration so that comparison to the exact truth can be made. We then investigate a moderate-dimensional example with 88 predictors (protein activity data). In this case, it is no longer possible to evaluate the exact posterior analytically. Instead, we apply an MCMC-based method to approximate the ``truth''---by running a very long chain. We compare the estimated PIPs produced by LIPS and BAS to these ``true'' values. (For problems of moderate dimensionality $p\leq 200$, the MCMC algorithm is able to produce reliable estimates of the PIPs, and this is validated in our examples by the almost identical estimates obtained under LIPS.) In addition, we compare the out-of-sample prediction of LIPS to those of three additional existing methods---BAS, iBMA, and LASSO. Our last two examples are based on simulation---they involve strongly correlated predictors and are of two different dimensionalities $p=100$ and $1,000$. For $p=100$, we again compare the results from our method to MCMC, BAS, iBMA, and LASSO. For the $p=1,000$ case, BAS cannot be applied and the MCMC approach has difficulty in exploring the vast model space effectively. We compare among LIPS, iBMA, and LASSO for prediction. The results show that LIPS is robust to both predictor correlation and increasing dimensionality.
For simplicity, in all the examples, we adopt the $g$-prior for the regression coefficients with the shrinkage parameter $g$ set to the number of observations.

\begin{exam}[US Crime data]
\label{ex:us_crime}
We first apply the LIPS algorithm to a classical data set introduced in \cite{vandaele:1978}.
It contains 15 variables and so an exhaustive computation of the marginal likelihood of all $2^{15}$ models is possible. Following \cite{clyde:2011}, we adopt a model space prior, called ``Beta-Binomial(1,1) prior'',  that assigns equal total probability---1/16---to each model size, and evenly splits the mass among models of each size.

In practice, for such low-dimensional problems one can enumerate the model space in little time, and so sampling-based methods are not really necessary. We use this example as a first demonstration of our method where truth is known exactly.

In particular, we adopt LIPS with $k=4$ and 5,000 particles to estimate the PIPs. To investigate the sampling distribution of the H-T estimators, we repeatedly applying LIPS 200 times. \ref{fig:us_crime} summarize the results, where we plot the mean and inter-quantile range of the each estimate over the 200 repeats. Note that each repeat is essentially an island of 5,000 particles, and so the average over the repeats gives an islanded estimate $\bar{\delta}_{HT}$. We mark the true PIPs computed through model space enumeration.
The (approximate) unbiasedness of the estimates is clearly demonstrated. In fact, each island of 5,000 particles alone provide a fairly reliable estimate of the PIPs (with small variances), while the islanded estimates essentially
recover the true values.
\begin{figure}[t]
\begin{center}
    \leavevmode
    \includegraphics[width=35em]{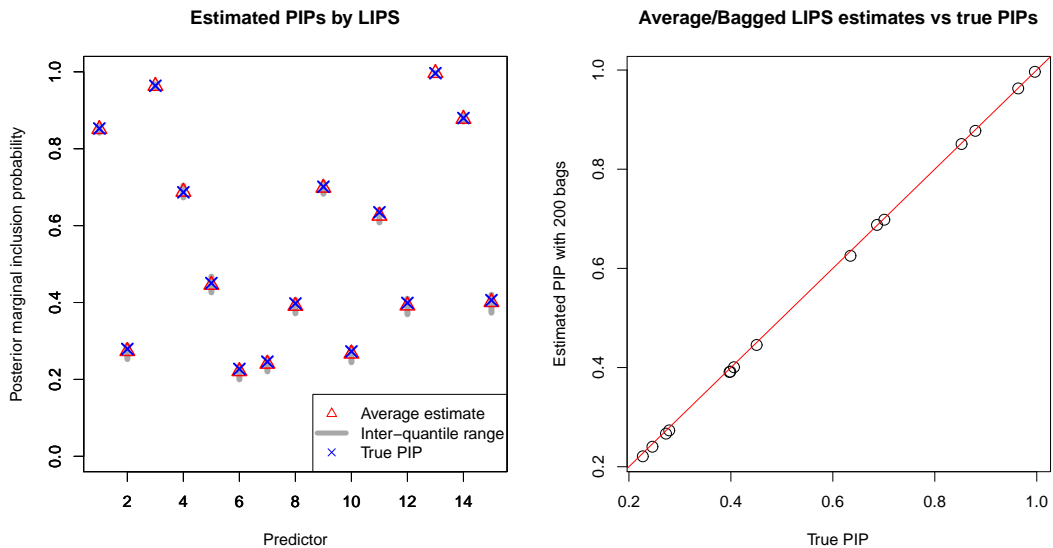}
\vspace{-1em}

     \caption{Left: Sampling distribution of the estimated PIPs by LIPS with $k=4$ and 5,000 particles for the US crime data. Gray band: The inter-quantile range of 200 estimates; red triangle: the islanded estimate---the average of 200 estimates; blue cross: true PIPs obtained through model space enumeration. Right: the islanded estimates for PIPs vs the true PIPs. The right line is the 45 degree line.}
\vspace{-1.5em}
  \label{fig:us_crime}
\end{center}
\end{figure}
\end{exam}

\begin{exam}[Protein activity data]
\label{ex:protein}
Our second example is a protein activity data set from \cite{clyde:1998}. Following those authors, we code the categorical variables by indicators and consider main effects and first-order interactions, resulting in a total of $p=88$ predictors. For problems with this many predictors, it is no longer possible to compute the exact model space posterior through enumeration. To get a handle on the truth as a baseline for comparison, we apply a popular MCMC-based method for BMA---Markov Chain Monte Carlo Model Composition, or MC$^3$, with add, delete, and swap moves \citep{madigan:1995,brown:1998,fernandez:2001} using the {\tt R} package {\tt BMS}. We estimate the PIPs by running a very long MC$^3$ chain ($2\times 10^7$ iterations with $1\times 10^7$ burn-in steps), and use these as proxies to the truth, to which we compare the estimates obtained from LIPS. (We will see that the results from LIPS almost match these proxies perfectly, confirming that these proxies are probably very close to the truth and so the MC$^3$ chain is long enough.)

We again adopt a Beta-Binomial(1,1) model space prior and apply LIPS with $k=3$. We generate 50,000 particles and repeat the computation 200 times, effectively creating 200 islands of particles. The sampling distribution of the 200 sets of esimates are given in \ref{fig:protein_pip} (left). The islanded estimates, that is the average of the 200 repeats, match the estimates from MC$^3$ very closely. The average number of steps before a particle stops is about 10, and thus the number of tentative models covered in each of the 200 LIPS run is about 500,000.

For comparison, we apply another method, BAS, to estimate the PIPs.
BAS samples models without replacement in an adaptive manner and carry out BMA based on the sampled models. Similarly, to get the sampling distribution of the BAS estimates, we repeatedly apply BAS (with $1\times 10^6$ model draws) 200 times, and compute the PIP estimates for each simulation. \ref{fig:protein_pip} (right) presents the results.

Comparing the results from LIPS and BAS, we find the two methods are complementary in several ways. First, LIPS gives unbiased esimates while BAS does not, but the LIPS estimates tend to be more variable than those from BAS. Moreover, the LIPS tends to provide most accurate estimates for PIPs that are very large or small, while giving the most variable estimates for PIPs that are close to 50\%. In contrast, BAS gives least biased estimates for PIPs that are close to 50\%, while it can substantially underestimate large PIPs and overestimate small PIPs. This observation suggests a potentially powerful hybrid algorithm that combine LIPS and BAS for estimating PIPs in moderate-dimensional problems. (The current implementation of BAS can only be applied to problems involving a few hundred predictors and $n>p+1$.)
\begin{figure}[tb]
\begin{center}
    \leavevmode
    \includegraphics[width=35em]{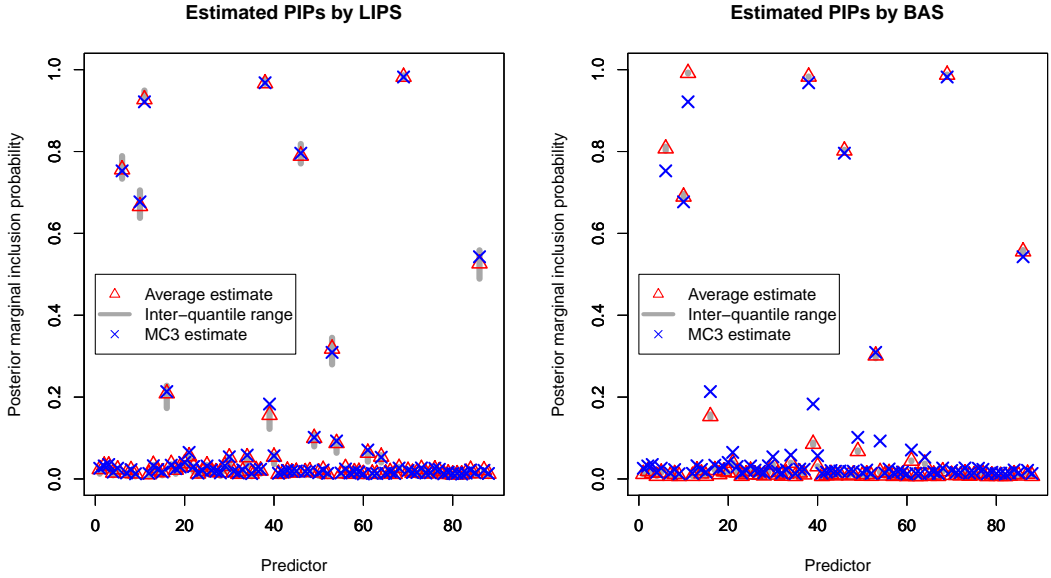}
\vspace{-1em}
     \caption{Left: Sampling distribution of the estimated PIPs by LIPS with $k=3$ and 50,000 particles for the protein activity data. Right: Sampling distribution of estimated PIPs by BAS with 1 million model draws. Gray band: the inter-quantile range of 200 estimates; red triangle: the average of 200 estimates; blue cross: PIPs estimated from a long MC$^{3}$ chain.}
  \label{fig:protein_pip}
\vspace{-1em}
\end{center}
\end{figure}

Next, we move onto out-of-sample prediction. To this end, we randomly split the data into a training set of $n_{train}=90$ observations and a testing set of $n_{test}=6$ observations. (BAS requires the training set to have a sample size $>p+1=89$.) We use the training set to get a model space posterior with which we predict the 6 testing observations using BMA. For comparison, we carry out the same prediction task using four methods---LIPS (with $k=3$ and 20,000 particles), BAS (with 1 million model draws), iBMA, and LASSO. For iBMA, we use the default setting that preserves a maximum of 30 predictors after iterated screening. For the LASSO, we use 10-fold cross-validation to select the shrinkage parameter.

We randomly divide the data into a training set and a testing set 200 times. 
For each random split, we apply the four methods, and to measure their performance, we compute the average squared error (ASE) for each method
\vspace{-1em}

\[
{\rm ASE}=\frac{1}{n_{test}}\sum_{j=1}^{n_{test}} (Y_{test,j}-\hat{Y}_{test,j})^2,
\]
\vspace{-2em}

\noindent where $\hat{Y}_{test,j}$ is the predicted value for the $j$th testing observation $Y_{test,j}$. In addition, we use the ASE averaged over the 200 random splits, denoted by $\overline{{\rm ASE}}$, as an overall performance metric for each method. \ref{fig:protein_ase} presents a scatter plot matrix of the ASEs for the four methods LIPS, BAS, iBMA, and LASSO (lower left triangle) as well as the pairwise differences in $\overline{{\rm ASE}}$ (upper right triangle). Overall, we see that the predictive performance of LIPS and that of BAS are very similar in this example---their difference is not statistically significant, and both perform substantially better than iBMA and LASSO for this example. We know that LIPS is unbiased but more variable than BAS, and so here this bias-variance trade-off about breaks even. For this example, LIPS, BAS, and iBMA all outperform LASSO by a substantial margin.

\begin{figure}[t]
\begin{center}
    \leavevmode
    \includegraphics[width=25em]{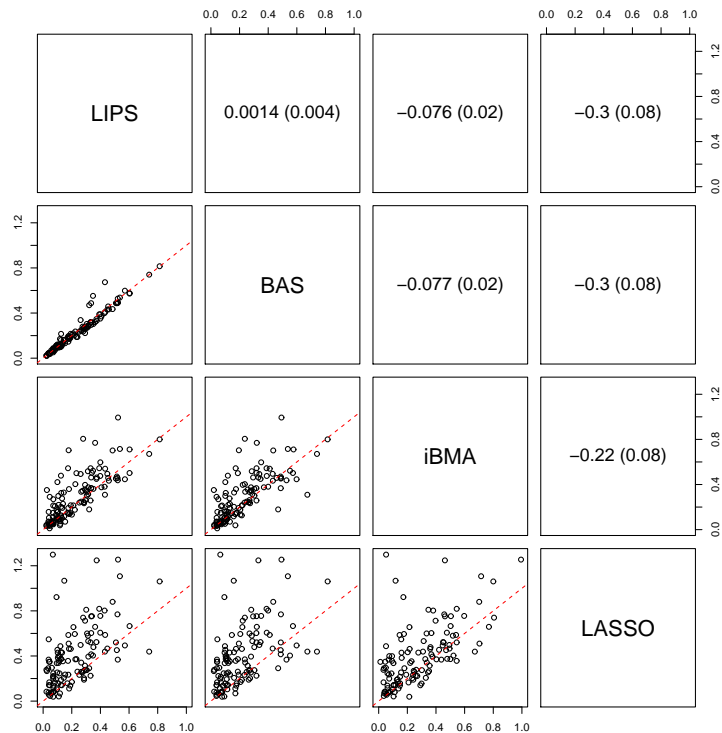}
    \vspace{-1em}

     \caption{Pairwise comparison of ASEs under 200 random splits of training and testing data for the protein activity example. Lower left triangle: Scatterplot matrix of ASEs for the four methods. The 45 degree lines are indicated by red dashes. Upper right triangle: The difference $\overline{{\rm ASE}}_{v} - \overline{{\rm ASE}}_{h}$ with the corresponding standard error given in parentheses, where $\overline{{\rm ASE}}_{v}$ is for the method on the vertical axis and $\overline{{\rm ASE}}_{h}$ the method on the horizontal axis.}
  \label{fig:protein_ase}
  \vspace{-1.5em}
\end{center}
\end{figure}

\end{exam}

\begin{exam}[Highly correlated predictors]
\label{ex:mod_dim}
In this example, we apply LIPS to a simulation scenario with a moderate number of dimensions ($p=100$) and the predictors are strongly correlated. The strong correlation makes the task of variable selection through estimating PIPs particularly challenging. Out-of-sample prediction, however, is not expected to be more difficult as one does not need to pin down the  ``causal'' predictors to get good predictions.

We simulate a data set of 350 observations each with 100 predictors, $X_1,X_2,\ldots,X_{100}$, and a response $Y$. The predictors are simulated from a multivariate normal distribution, with marginal means all equal to 0 and marginal variances all equal to 1, along with the following correlation structure
$
\corr(X_i,X_j) = (1 - 0.05|i-j|)\I_{|i-j|\leq 20}.
$
So ``close-by'' predictors are highly correlated and the correlation decays with their ``distance''. Such a local correlation structure is common in applications. For example, in genomic studies where the predictors are gene markers, nearby markers on a chromosome are generally strongly correlated.

The response $Y$ is simulated according to the following model
\vspace{-1.5em}

\[
Y = 10 + 3 X_2 -3 X_{30} + 3 X_{58} -3 X_{75} + 3 X_{97} + \epsilon
\]
\vspace{-2.5em}

\noindent where the errors are independent draws from a Normal($0,10^2$) distribution. We again adopt a Beta-Binomial(1,1) model space prior that places equal probability, 1/101, to each model size ranging from 0 to 100.

Again, we apply LIPS with $k=3$ and 50,000 particles as well as BAS with 1 million model draws to estimate the PIPs, and compare the estimates to those obtained from a very long MC$^3$ chain with add, delete, and swap steps ($2\times 10^7$ iterations with $1\times 10^7$ burn-in steps). The average number of steps before a particle stops is about 6.4, and thus the number of tentative models covered in each of the 200 LIPS run is about 320,000. The results are presented in \ref{fig:mod_dim}. Similar to the earlier examples, LIPS based H-T estimators are (approximately) unbiased and the average of the 200 repeats, which are islanded estimates, match the MC$^{3}$ almost exactly, thereby confirming the validity of each other. On the other hand, the BAS estimates tend to overestimate small PIPs and underestimate large PIPs. In contrast to the protein example, the BAS estimates now demonstrate more variability. This is a direct consequence of the strong local correlation among the predictors---due to the competition among neighboring predictors, BAS becomes less aggressive in choosing the models to be sampled.

\begin{figure}[t]

\begin{center}
    \leavevmode
    \includegraphics[width=35em]{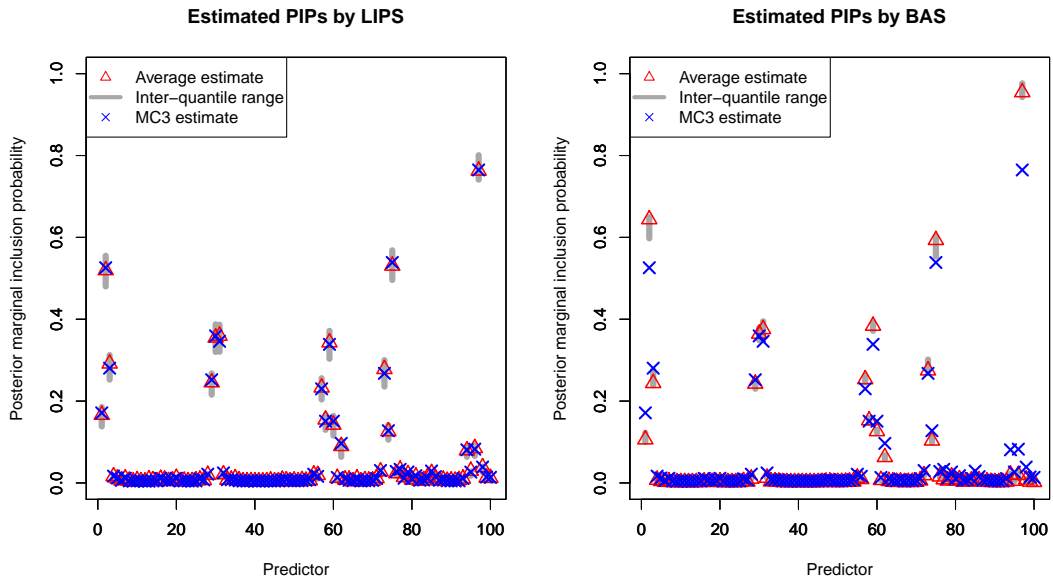}
\vspace{-1em}

     \caption{Left: Sampling distribution of the estimated PIPs by LIPS with $k=3$ and 50,000 particles for Example~\ref{ex:mod_dim}. Right: Sampling distribution of the estimated PIPs by BAS with 1 million model draws. Dark gray band: the inter-quantile range of the 200 estimates; red triangle: the islanded estimate---the average of the 200 estimates; blue cross: PIPs estimated from a long MC$^3$ chain with add, delete, and swap steps.}
  \label{fig:mod_dim}
  \vspace{-1em}
\end{center}
\end{figure}

Next we again compare the out-of-sample prediction performance of LIPS, BAS, iBMA, and LASSO. We repeatedly simulate training and testing sets with $n_{train}=200$ and $n_{test}=150$ and use ASE to measure the performance of the methods. \ref{fig:mod_dim_new_pred_ase} presents a scatter plot matrix of the ASEs (lower left triangle) as well as the pairwise differences in the overall average ASEs (upper right triangle). The predictive performance of all four methods are similar---the ASEs generally lie around the 45 degree lines in the pairwise scatter plots. Their performance difference is relatively small but statistically significant. Specifically, LIPS and LASSO achieved the best overall average ASE, followed by BAS and iBMA.

\begin{figure}[t]
\begin{center}
    \leavevmode
    \includegraphics[width=25em]{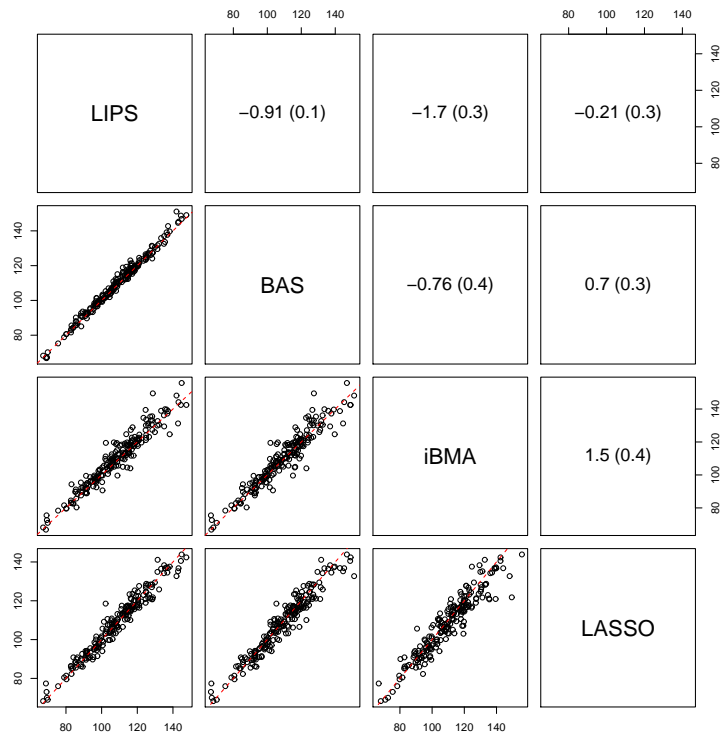}
\vspace{-1em}
     \caption{Pairwise comparison of ASEs under 200 simulations of training and testing data for Example~\ref{ex:mod_dim}. Lower left triangle: Scatterplot matrix of ASEs for the four methods. The 45 degree lines are indicated by red dashes. Upper right triangle: The difference $\overline{{\rm ASE}}_{v} - \overline{{\rm ASE}}_{h}$ with the corresponding standard error given in parentheses, where $\overline{{\rm ASE}}_{v}$ is for the method on the vertical axis and $\overline{{\rm ASE}}_{h}$ the method on the horizontal axis.}
  \label{fig:mod_dim_new_pred_ase}
\vspace{-1.5em}
\end{center}
\end{figure}
\end{exam}

Up till this point, our examples are of low enough dimension that either allows model space enumeration or effective exploration through MCMC algorithms. In the final example, we move onto a high-dimensional example involving 1,000 predictors while maintaining the high local correlation structure among the predictors as in the previous example.
\begin{exam}[High-dimensional model space]
\label{ex:hi_dim}
We consider a high-dimensional version of the previous example---a scenario with 1,000 potential predictors, $X_1$, $X_2$,\ldots, $X_{1000}$.
The predictors are again multivariate normal, with marginal means all equal to 0 and marginal variances all equal to 1, along with the same correlation structure as in the previous example
$
\corr(X_i,X_j) = (1 - 0.05|i-j|)\I_{|i-j|\leq 20}.
$
A response $Y$ is simulated as
\vspace{-1em}

\[
Y = 10 + 3 X_{120} - 3 X_{280} + 3 X_{400} -  3 X_{560} + 3 X_{807} + \epsilon
\]
\vspace{-2em}

\noindent where the errors are independent draws from a Normal($0,10^2$) distribution. We simulate such data sets of size 700. We adopt a symmetric model space prior that puts equal prior probability to each model size from 0 to 100.

As before, to evaluate the sampling properties of the estimates we carry out 200 runs of LIPS with $k=2$ and 20,000 particles. The average number of steps before a particle stops is about 7.4, and thus on average the number of tentative models covered in each of the 200 LIPS run is about 148,000. \ref{fig:high_dim} presents the sampling distribution of the estimated PIPs. The computation for each run takes about 30 hours on a single 3.6GHz CPU core and 400 Mbs of RAM. Aside from the larger variance, the behavior of the H-T estimates look qualitatively similar to those in the 100-dimensional example, which is reassuring as the two examples are simulated under the same correlation structure and similar response models.
\begin{figure}[bt]
\begin{center}
    \leavevmode
    \includegraphics[width=32em]{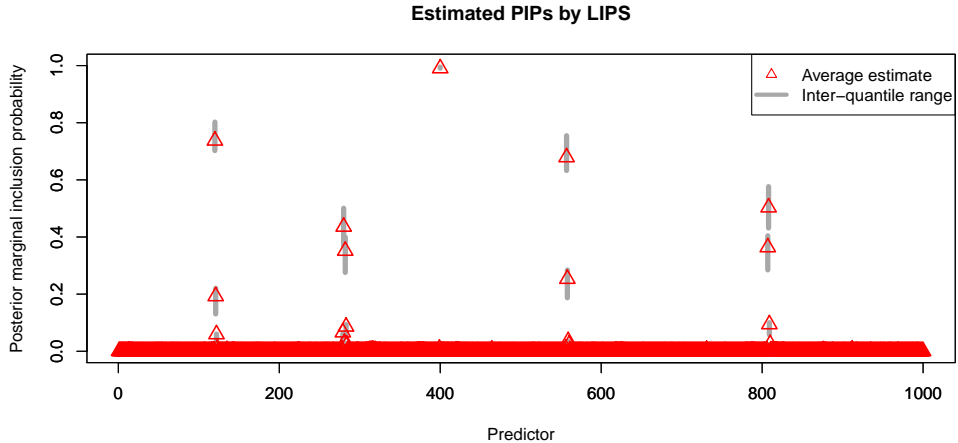}
    \vspace{-1em}
     \caption{Sampling distribution of estimated PIPs by LIPS with $k=2$ and 20,000 particles in the high-dimensional example. Gray band: inter-quantile of 200 estimates; red triangle: the islanded estimate---the average of 200 estimates. The MC$^3$ with add, delete, and swap fails to produce meaningful estimates after 100 million iterations with 50 million burn-ins and so the corresponding results are not plotted.}
  \label{fig:high_dim}
\vspace{-1em}
\end{center}
\end{figure}

The version of the {\tt BAS} package in {\tt R} cannot be run on problems with $p\geq n$ or $p\geq 1000$.
We have tried MC$^3$ with add, delete, and swap moves on this example using the {\tt R} package {\tt BMS} as we did for the 100-dimensional example. However we could not get the chain to converge within reasonable time. With 100 million iterations and 50 million burn-in steps or about 5 days of running time on a single Intel Core-i7 3.6GHz CPU core, the estimated PIPs from the MCMC run still do not make sense and so are omitted. We do acknowledge that someone more experienced with MCMC will be able to construct a more effective MCMC that can converge and mix better, possibly using population-based strategies as those developed in \cite{liang:2000,jasra:2007,bottolo:2010}.

\begin{figure}[t]
\begin{center}
    \leavevmode
    \includegraphics[width=25em]{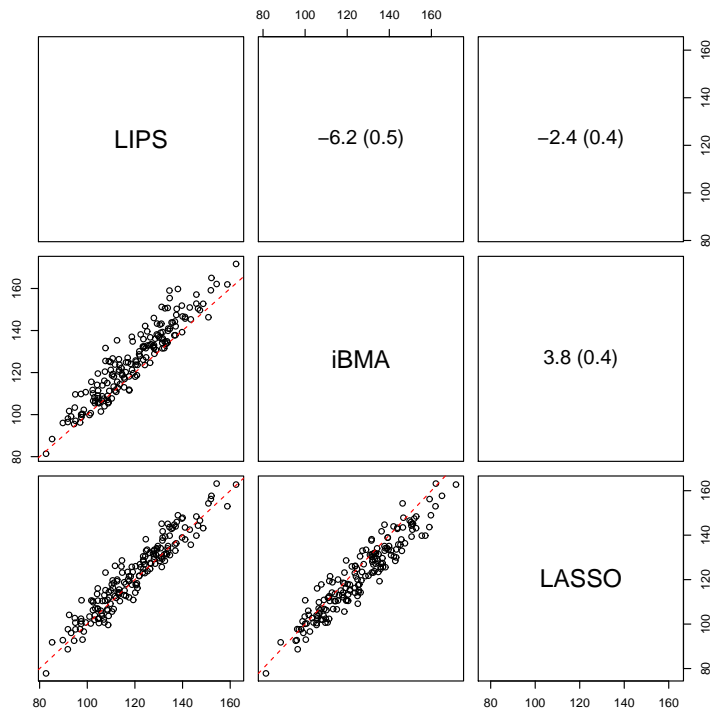}
    \vspace{-1em}
     \caption{Pairwise comparison of ASEs under 200 simulations of training and testing data for the high dimensional example. Lower left triangle: Scatterplot matrix of ASEs for the four methods. The 45 degree lines are indicated by red dashes. Upper right triangle: The difference $\overline{{\rm ASE}}_{v} - \overline{{\rm ASE}}_{h}$ with the corresponding standard error given in parentheses, where $\overline{{\rm ASE}}_{v}$ is for the method on the vertical axis and $\overline{{\rm ASE}}_{h}$ the method on the horizontal axis.}
  \label{fig:high_dim_pred_ase}
  \vspace{-1.5em}
\end{center}
\end{figure}

We again consider the problem of out-of-sample prediction and compare LIPS to iBMA and LASSO. We randomly simulate a training and a testing set with $n_{train}=600$ and $n_{test}=100$ for 200 times. For each simulated training and testing pair, we apply LIPS (with $k=2$ and 20,000 particles), iBMA, and LASSO (with 10-fold CV). \ref{fig:high_dim_pred_ase} presents the pairwise comparison of the ASEs. LIPS shows the best predictive performance among the three, followed by LASSO, and then iBMA. While we show the results here for LIPS with 20,000 particles, interestingly we found that in this example LIPS with even just 200 particles outperforms the other two methods in predictive performance. On the other hand, with so few particles, the models that LIPS samples become very variable due to the high predictor correlation, and so the corresponding PIP estimates are unreliable. This is exactly what we expect---the strong predictor correlation makes estimating PIPs harder but not out-of-sample prediction.
\end{exam}

\vspace{-2.2em}

\section{Discussion}
\label{sec:discussion}
\vspace{-0.7em}

In this work we have showed that one can use a probabilistic version of the classical forward-stepwise variable selection procedure as a data-augmentation scheme for model space distributions, and that due to the Markov property of this representation it allows us to use information propagation methods to achieve scalable posterior sampling on model spaces.

It is worth noting some similarities and differences between LIPS and MCMC. LIPS propagates each particle---a sequence of tentative models---through a Markov chain, and thus is operationally similar to an MCMC algorithm, especially a multi-chain MCMC algorithm. An important difference exists in the purpose of the Markov chain in each of these algorithms. In LIPS the Markov chain is a latent variable representation for the target distribution, from which independent samples are drawn, whereas in MCMC, the Markov Chain has the posterior as the stationary distribution and dependent samples are generate from the chain. 
Another difference between LIPS and MCMC lies in the mechanism that determines the convergence of the two methods. LIPS requires a large number of short chains to obtain reliable estimates, and thus is easily parallelizable. In contrast, MCMC relies on Markov chain convergence, and thus requires long chains to achieve convergence. Multi-chain MCMC typically involves a {\em small} number of very {\em long} chains, and their purpose is to help effectively explore the topological structure of the often multi-modal posterior.

A key idea in LIPS is that by finding a highly effective proposal through information propagation, one can achieve very reliable estimates with a relatively small number of particles. In our numerical examples, for instance, tens of thousands of particles appear to be sufficient. 
The number of particles in LIPS (and more generally the sample size in importance sampling) for achieving desired accuracy depends critically on the effectiveness of the proposals. If the proposal distribution is far from the actual posterior, then a very large number of particles will be necessary.

The efficiency of the proposed LIPS algorithm also depends on the sparsity of the underlying model. Monte Carlo errors accumulate through the stepwise inclusion, and so if the true model actually involves a large number of covariates, it will be hard to reach the regions in the model space near the true model through stochastic stepwise inclusion. In such cases, incorporating a resampling step \citep{liu:2001} can help remedy (though not eliminate) the problem. 

In classical stepwise variable selection, incorporating a backward removal step can often improve the model selected. One may thus consider building a probabilistic version of a bi-directional stepwise procedure to further improve sampling efficiency. To extend the pFS representation to a bi-directional one, a few technical complications need to be addressed and this is currently under investigation. For example, because such a bi-directional chain does not necessarily terminate in finite steps, the prior probability over model size under the bi-directional chain can no longer be computed under a simple finite recursion formula. In this case, a potential link to birth-death process can be very useful for prior specification and posterior computation under the bi-directional augmentation. A bi-directional augmentation could also improve the samples efficiency for problems involving a relatively large number of predictors. In this case, one could set the initial model to be some baseline model that contains the predictors obtained using a simple variable selection algorithm such as the LASSO. The LIPS chain can start from the baseline propose in the nearby model space, resulting in shorter chains.

In some problems model selection may be more useful than prediction. For example, in genetic studies where the interest is in identifying the relevant gene markers associated with a phenotype, a single model is often desired for its interpretability.
To this end, one can use BMA (and thus LIPS) for model selection. One effective approach is the so-called ``median probability model'' method \citep{scott:2010}, which is to select the model that contains the variables whose posterior inclusion probabilities are more than 50\%.

In these variable selection problems, the PIP computed under BMA can serve as a Bayesian ``$p$-value'' that provides a probabilistic summary of one's confidence in the inclusion of each predictor. Moreover, one can compute quantities such as posterior {\em co-inclusion} probabilities for groups of variables, e.g.\ those that fall in the same biological pathway. More specifically, one can estimate the posterior probability for events such as (i) at least one variable in the group is included and (ii) all variables in the group are included, and use this probability as a measure of significance for the group as a whole.

Finally, one of the main uses of the pFS representation is for constructing effective model space proposal distributions through LIP. It is thus natural to consider adopting this Markov augmentation and LIP idea for constructing effective MCMC proposals. This is currently under investigation.

\section*{Acknowledgment}

This research is supported by NSF grant DMS-1309057.
The author is especially grateful to Quanli Wang for help in programming that substantially improved the efficiency of the software. The implementation of the LIPS algorithm used in the examples is based on the {\tt SMCTC} template class in {\tt C++} \citep{johansen:2009}.

\bibliography{ModelSel}

\begin{thebibliography}{}

\bibitem[\protect\citeauthoryear{Annest, Bumgarner, Raftery, and Yeung}{Annest
  et~al.}{2009}]{annest:2009}
Annest, A., R.~E. Bumgarner, A.~E. Raftery, and K.~Y. Yeung (2009).
\newblock Iterative bayesian model averaging: a method for the application of
  survival analysis to high-dimensional microarray data.
\newblock {\em {BMC Bioinformatics}\/}~{\em 10\/}(72).

\bibitem[\protect\citeauthoryear{Berger and Molina}{Berger and
  Molina}{2005}]{berger:2005}
Berger, J.~O. and G.~Molina (2005).
\newblock Posterior model probabilities via path-based pairwise priors.
\newblock {\em Statistica Neerlandica\/}~{\em 59\/}(1), 3--15.

\bibitem[\protect\citeauthoryear{Bottolo and Richardson}{Bottolo and
  Richardson}{2010}]{bottolo:2010}
Bottolo, L. and S.~Richardson (2010).
\newblock {Evolutionary Stochastic Search for {B}ayesian model exploration}.
\newblock {\em {B}ayesian {A}nalysis\/}~{\em 5\/}(3), 583--618.

\bibitem[\protect\citeauthoryear{Brown, Vannucci, and Fearn}{Brown
  et~al.}{1998}]{brown:1998}
Brown, P.~J., M.~Vannucci, and T.~Fearn (1998).
\newblock Multivariate {B}ayesian variable selection and prediction.
\newblock {\em Journal of the Royal Statistical Society: Series B\/}~{\em
  60\/}(3), 627--641.

\bibitem[\protect\citeauthoryear{Capp\'{e}, Godsill, and Moulines}{Capp\'{e}
  et~al.}{2007}]{cappe:2007}
Capp\'{e}, O., S.~J. Godsill, and E.~Moulines (2007, May).
\newblock {An Overview of Existing Methods and Recent Advances in Sequential
  Monte Carlo}.
\newblock {\em Proceedings of the IEEE\/}~{\em 95\/}(5), 899--924.

\bibitem[\protect\citeauthoryear{Chopin}{Chopin}{2002}]{chopin:2002}
Chopin, N. (2002).
\newblock A sequential particle filter method for static models.
\newblock {\em Biometrika\/}~{\em 89\/}(3), 539--552.

\bibitem[\protect\citeauthoryear{Clyde and Ghosh}{Clyde and
  Ghosh}{2012}]{clyde:2012}
Clyde, M.~A. and J.~Ghosh (2012).
\newblock A note on the bias in estimating posterior probabilities in variable
  selection.
\newblock Technical report, Duke University.

\bibitem[\protect\citeauthoryear{Clyde, Ghosh, and Littman}{Clyde
  et~al.}{2011}]{clyde:2011}
Clyde, M.~A., J.~Ghosh, and M.~L. Littman (2011).
\newblock {B}ayesian adaptive sampling for variable selection and model
  averaging.
\newblock {\em Journal of Computational and Graphical Statistics\/}~{\em
  20\/}(1), 80--101.

\bibitem[\protect\citeauthoryear{Clyde and Parmigiani}{Clyde and
  Parmigiani}{1998}]{clyde:1998}
Clyde, M.~A. and G.~Parmigiani (1998).
\newblock Protein construct storage: {B}ayesian variable selection and
  prediction with mixtures.
\newblock {\em Journal of Biopharmaceutical Statistics\/}~{\em 8\/}(3),
  431--443.

\bibitem[\protect\citeauthoryear{Del~Moral, Doucet, and Jasra}{Del~Moral
  et~al.}{2006}]{delmoral:2006}
Del~Moral, P., A.~Doucet, and A.~Jasra (2006).
\newblock Sequential monte carlo samplers.
\newblock {\em Journal of the Royal Statistical Society: Series B (Statistical
  Methodology)\/}~{\em 68\/}(3), 411--436.

\bibitem[\protect\citeauthoryear{Fern\'andez, Ley, and Steel}{Fern\'andez
  et~al.}{2001}]{fernandez:2001}
Fern\'andez, C., E.~Ley, and M.~F.~J. Steel (2001).
\newblock Benchmark priors for {B}ayesian model averaging.
\newblock {\em Journal of Econometrics\/}~{\em 100}, 381--427.

\bibitem[\protect\citeauthoryear{Gelman and Meng}{Gelman and
  Meng}{1998}]{gelman:1998}
Gelman, A. and X.-L. Meng (1998).
\newblock Simulating normalizing constants: from importance sampling to bridge
  sampling to path sampling.
\newblock {\em Statistical Science\/}~{\em 13\/}(2), 163--185.

\bibitem[\protect\citeauthoryear{George}{George}{1999}]{george:1999}
George, E.~I. (1999).
\newblock Sampling considerations for model averaging and model search. invited
  discussion of ``{M}odel averaging and model search'', by {M}. {C}lyde.
\newblock In J.~M. Bernado, J.~O. Berger, A.~P. Dawid, and A.~F.~M. Smith
  (Eds.), {\em {B}ayesian Statistics 6}, pp.\  175--177. Oxford, UK: Oxford
  University Press.

\bibitem[\protect\citeauthoryear{George}{George}{2010}]{george:2010}
George, E.~I. (2010).
\newblock Dilution priors: {C}ompensating for model space redundancy.
\newblock In {\em Borrowing Strength: Theory Powering Applications - A
  Festschrift for Lawrence Brown}, pp.\  158--165. IMS Collections.

\bibitem[\protect\citeauthoryear{George and McCulloch}{George and
  McCulloch}{1993}]{george:1993}
George, E.~I. and R.~E. McCulloch (1993).
\newblock {Variable Selection Via Gibbs Sampling}.
\newblock {\em Journal of the American Statistical Association\/}~{\em
  88\/}(423), 881--889.

\bibitem[\protect\citeauthoryear{George and McCulloch}{George and
  McCulloch}{1997}]{george:1997}
George, E.~I. and R.~E. McCulloch (1997).
\newblock Approaches for {B}ayesian variable selection.
\newblock {\em Statistica Sinica\/}~{\em 7\/}(2), 339--373.

\bibitem[\protect\citeauthoryear{Geweke}{Geweke}{1996}]{geweke:1996}
Geweke, J. (1996).
\newblock Variable selection and model comparison in regression.
\newblock In J.~M. Bernado, J.~O. Berger, A.~P. Dawid, and A.~F.~M. Smith
  (Eds.), {\em {B}ayesian Statistics 5}, pp.\  339--348. Oxford, UK: Oxford
  University Press.

\bibitem[\protect\citeauthoryear{Ghosh and Clyde}{Ghosh and
  Clyde}{2011}]{ghosh:2011}
Ghosh, J. and M.~A. Clyde (2011).
\newblock Rao–blackwellization for bayesian variable selection and model
  averaging in linear and binary regression: A novel data augmentation
  approach.
\newblock {\em Journal of the American Statistical Association\/}~{\em
  106\/}(495), 1041--1052.

\bibitem[\protect\citeauthoryear{Hans, Dobra, and West}{Hans
  et~al.}{2007}]{hans:2007}
Hans, C., A.~Dobra, and M.~West (2007, June).
\newblock {Shotgun Stochastic Search for ``Large p'' Regression}.
\newblock {\em Journal of the American Statistical Association\/}~{\em
  102\/}(478), 507--516.

\bibitem[\protect\citeauthoryear{Heaton and Scott}{Heaton and
  Scott}{2010}]{heaton:2010}
Heaton, M.~J. and J.~G. Scott (2010).
\newblock {B}ayesian computation and the linear model.
\newblock In M.-H. Chen, D.~Dey, P.~Mueller, and D.~Sun (Eds.), {\em Frontiers
  of Statistical Decision Making and {B}ayesian Analysis}, pp.\  527--545. New
  York: Springer-Verlag.

\bibitem[\protect\citeauthoryear{Hoeting, Madigan, Raftery, and
  Volinsky}{Hoeting et~al.}{1999}]{hoeting:1999}
Hoeting, J.~A., D.~Madigan, A.~E. Raftery, and C.~T. Volinsky (1999).
\newblock {B}ayesian model averaging: A tutorial.
\newblock {\em Statistical Science\/}~{\em 14\/}(4), pp. 382--401.

\bibitem[\protect\citeauthoryear{Horvitz and Thompson}{Horvitz and
  Thompson}{1952}]{horvitz:1952}
Horvitz, D.~G. and D.~J. Thompson (1952).
\newblock A generalization of sampling without replacement from a finite
  universe.
\newblock {\em Journal of the American Statistical Association\/}~{\em
  47\/}(260), pp. 663--685.

\bibitem[\protect\citeauthoryear{Jasra, Stephens, and Holmes}{Jasra
  et~al.}{2007}]{jasra:2007}
Jasra, A., D.~A. Stephens, and C.~C. Holmes (2007).
\newblock Population-based reversible jump markov chain monte carlo.
\newblock {\em Biometrika\/}~{\em 94\/}(4), 787--807.

\bibitem[\protect\citeauthoryear{Johansen}{Johansen}{2009}]{johansen:2009}
Johansen, A.~M. (2009).
\newblock Smctc: Sequential monte carlo in {C}++.
\newblock {\em Journal of Statistical Software\/}~{\em 30\/}(6), 1--41.

\bibitem[\protect\citeauthoryear{Jones, Carvalho, Dobra, Hans, Carter, and
  West}{Jones et~al.}{2005}]{jones:2005}
Jones, B., C.~Carvalho, A.~Dobra, C.~Hans, C.~Carter, and M.~West (2005).
\newblock Experiments in stochastic computation for high-dimensional graphical
  models.
\newblock {\em Statistical Science\/}~{\em 20\/}(4), 388--400.

\bibitem[\protect\citeauthoryear{Lakshminarayanan, Roy, and
  Teh}{Lakshminarayanan et~al.}{2013}]{lak:2013}
Lakshminarayanan, B., D.~Roy, and Y.~W. Teh (2013).
\newblock Top-down particle filtering for {B}ayesian decision trees.
\newblock In {\em ICML}.

\bibitem[\protect\citeauthoryear{Liang, Paulo, Molina, Clyde, and Berger}{Liang
  et~al.}{2008}]{liang:2008}
Liang, F., R.~Paulo, G.~Molina, M.~A. Clyde, and J.~O. Berger (2008).
\newblock {Mixtures of g-Priors for {B}ayesian Variable Selection}.
\newblock {\em Journal of the American Statistical Association\/}~{\em
  103\/}(481), 410--423.

\bibitem[\protect\citeauthoryear{Liang and Wong}{Liang and
  Wong}{2000}]{liang:2000}
Liang, F. and W.~H. Wong (2000).
\newblock Evolutionary {M}onte {C}arlo: Applications to {$C_p$} model sampling
  and change point problem.
\newblock {\em Statistica Sinica\/}~{\em 10}, 317--342.

\bibitem[\protect\citeauthoryear{Lin, Chen, and Liu}{Lin
  et~al.}{2013}]{lin:2013}
Lin, M., R.~Chen, and J.~S. Liu (2013, February).
\newblock {Lookahead Strategies for Sequential Monte Carlo}.
\newblock {\em Statistical Science\/}~{\em 28\/}(1), 69--94.

\bibitem[\protect\citeauthoryear{Liu}{Liu}{2001}]{liu:2001}
Liu, J.~S. (2001, January).
\newblock {\em {Monte Carlo Strategies in Scientific Computing}}.
\newblock Springer.

\bibitem[\protect\citeauthoryear{Madigan, York, and Allard}{Madigan
  et~al.}{1995}]{madigan:1995}
Madigan, D., J.~York, and D.~Allard (1995).
\newblock {{B}ayesian Graphical Models for Discrete Data}.
\newblock {\em International Statistical Review / Revue Internationale de
  Statistique\/}~{\em 63\/}(2), 215--232.

\bibitem[\protect\citeauthoryear{Nott and Green}{Nott and
  Green}{2004}]{nott:2004}
Nott, D.~J. and P.~J. Green (2004).
\newblock {B}ayesian variable selection and the swendsen-wang algorithm.
\newblock {\em Journal of Computational and Graphical Statistics\/}~{\em
  13\/}(1), pp. 141--157.

\bibitem[\protect\citeauthoryear{Nott and Kohn}{Nott and
  Kohn}{2005}]{nott:2005}
Nott, D.~J. and R.~Kohn (2005).
\newblock Adaptive sampling for bayesian variable selection.
\newblock {\em Biometrika\/}~{\em 92\/}(4), 747--763.

\bibitem[\protect\citeauthoryear{Raftery and Zheng}{Raftery and
  Zheng}{2003}]{raftery:2003}
Raftery, A. and Y.~Zheng (2003).
\newblock Discussion: Performance of {B}ayesian model averaging.
\newblock {\em Journal of the American Statistical Association\/}~{\em 98},
  931--938.

\bibitem[\protect\citeauthoryear{Rockova and George}{Rockova and
  George}{2014}]{rockova:2014}
Rockova, V. and E.~I. George (2014).
\newblock Emvs: The em approach to bayesian variable selection.
\newblock {\em Journal of the American Statistical Association\/}~{\em
  109\/}(506), 828--846.

\bibitem[\protect\citeauthoryear{Sch\"afer and Chopin}{Sch\"afer and
  Chopin}{2013}]{schaffer:2013}
Sch\"afer, C. and N.~Chopin (2013).
\newblock Sequential {M}onte {C}arlo on large binary sampling spaces.
\newblock {\em Statistics and Computing\/}~{\em 23\/}(2), 163--184.

\bibitem[\protect\citeauthoryear{Scott and Berger}{Scott and
  Berger}{2010}]{scott:2010}
Scott, J.~G. and J.~O. Berger (2010).
\newblock {B}ayes and empirical-{B}ayes multiplicity adjustment in the
  variable-selection problem.
\newblock {\em Annals of Statistics\/}~{\em 38}, 2587.

\bibitem[\protect\citeauthoryear{Scott and Carvalho}{Scott and
  Carvalho}{2008}]{scott:2008}
Scott, J.~G. and C.~M. Carvalho (2008).
\newblock Feature-inclusion stochastic search for gaussian graphical models.
\newblock {\em Journal of Computational and Graphical Statistics\/}~{\em
  17\/}(4), 790--808.

\bibitem[\protect\citeauthoryear{Shi and Dunson}{Shi and
  Dunson}{2011}]{shi:2011}
Shi, M. and D.~B. Dunson (2011).
\newblock {B}ayesian variable selection via particle stochastic search.
\newblock {\em {S}tatistics \& {P}robability {L}etters\/}~{\em 81\/}(2),
  283--291.

\bibitem[\protect\citeauthoryear{Smith and Kohn}{Smith and
  Kohn}{1996}]{smith:1996}
Smith, M. and R.~Kohn (1996).
\newblock Nonparametric regression using {B}ayesian variable selection.
\newblock {\em Journal of Econometrics\/}~{\em 75\/}(2), 317 -- 343.

\bibitem[\protect\citeauthoryear{Tibshirani}{Tibshirani}{2011}]{tibs:2011}
Tibshirani, R. (2011, June).
\newblock {Regression shrinkage and selection via the lasso: a retrospective}.
\newblock {\em Journal of the Royal Statistical Society: Series B (Statistical
  Methodology)\/}~{\em 73\/}(3), 273--282.

\bibitem[\protect\citeauthoryear{Vandaele}{Vandaele}{1978}]{vandaele:1978}
Vandaele, W. (1978).
\newblock Participation in illegitimate activities---{E}hrlich revisited.
\newblock In A.~Blumstein, J.~Cohen, and D.~Nagin (Eds.), {\em Deterrence and
  Incapacitation}, pp.\  270--335. Washington, DC: National Academy of Sciences
  Press.

\bibitem[\protect\citeauthoryear{Wilson, Iversen, Clyde, Schmidler, and
  Schildkraut}{Wilson et~al.}{2010}]{wilson:2010}
Wilson, M.~A., E.~S. Iversen, M.~A. Clyde, S.~C. Schmidler, and J.~M.
  Schildkraut (2010).
\newblock Bayesian model search and multilevel inference for snp association
  studies.
\newblock {\em The Annals of Applied Statistics\/}~{\em 4\/}(3), 1342--1364.

\bibitem[\protect\citeauthoryear{Zellner}{Zellner}{1981}]{zellner:1981}
Zellner, A. (1981).
\newblock Posterior odds ratios for regression hypotheses: General
  considerations and some specific results.
\newblock {\em Journal of Econometrics\/}~{\em 16\/}(1), 151--152.

\end{thebibliography}


\begin{thebibliography}{}

\bibitem[\protect\citeauthoryear{Berger and Pericchi}{Berger and
  Pericchi}{2001}]{berger:2001}
Berger, J.~O. and L.~R. Pericchi (2001).
\newblock Objective {B}ayesian methods for model selection: Introduction and
  comparison.
\newblock {\em Lecture Notes-Monograph Series\/}~{\em 38}, pp. 135--207.

\bibitem[\protect\citeauthoryear{George}{George}{1999}]{george:1999}
George, E.~I. (1999).
\newblock Sampling considerations for model averaging and model search. invited
  discussion of ``{M}odel averaging and model search'', by {M}. {C}lyde.
\newblock In J.~M. Bernado, J.~O. Berger, A.~P. Dawid, and A.~F.~M. Smith
  (Eds.), {\em {B}ayesian Statistics 6}, pp.\  175--177. Oxford, UK: Oxford
  University Press.

\bibitem[\protect\citeauthoryear{Liang, Paulo, Molina, Clyde, and Berger}{Liang
  et~al.}{2008}]{liang:2008}
Liang, F., R.~Paulo, G.~Molina, M.~A. Clyde, and J.~O. Berger (2008).
\newblock {Mixtures of g-Priors for {B}ayesian Variable Selection}.
\newblock {\em Journal of the American Statistical Association\/}~{\em
  103\/}(481), 410--423.

\end{thebibliography}

\end{document}


\maketitle

\doublespace

\section*{S1. Proofs}

\begin{proof}[Proof of Theorem~1]
The result follows immediately from the distributions of the decision variables and the fact that the pFS procedure stops in the first $t-1$ steps iff $|\gamma_{(t-1)}| < t-1$.
\end{proof}

\begin{proof}[Proof of Theorem~2]
Our proof strategy is to actually find a pFS representation for any model space distribution $\pi$. To this end, we can proceed by induction on the total number of potential predictors. First, we note that the conclusion holds for $p=1$ or $\om_{\M}=\{0,1\}$. In this case there are but two models in the space: the null model and the model including $\bX_1$, written as $(0)$ and $(1)$ respectively. Let $\pi(\cdot)$ be any probability distribution on $\om_{\M}$. It is easy to check that $\pi(\cdot)$ is the marginal distribution of the final model under the pFS procedure with $\rho(0)=\pi(0)$, $\rho(1)=1$, and $\lambda_1(0)=1$.

Now suppose the inductive claim holds for any model space involving up to $p-1$ variables. We next show it must hold for the one with $p$ predictors, or $\om_{\M}=\{0,1\}^{p}$, as well. To this end, again let $\pi(\cdot)$ be any distribution on $\{0,1\}^{p}$, and let $\om_{\M}^{(-p)}=\{0,1\}^{p-1}\times \{0\}$---the collection of models that do not involve $\bX_p$. Let us define a new distribution $\pi'(\cdot)$ on $\om_{\M}^{(-p)}$  such that for each $\gamma\in\om_{\M}^{(-p)}$,
\[
\pi'(\gamma) = \pi(\gamma) + \pi(\gamma^{+p}).
\]
where $\gamma^{+p}\in \om_{\M}$ is the model that adds an additional variable, $\bX_p$, into $\gamma$. It is easy to check that $\sum_{\gamma\in \om^{(-p)}_{\M}}\pi'(\gamma)=1$.

Because $\om_{\M}^{(-p)}$ is ``isomorphic'' to $\{0,1\}^{p-1}$,
$\pi'(\cdot)$ can be considered a probability distribution on $\{0,1\}^{p-1}$.
Thus by the inductive hypothesis, $\pi'$ has a pFS representation with parameter mappings $\rho'$ and $\blam'$ defined on $\{0,1\}^{p-1}\backslash\{1,1,\ldots,1\}$.
Now for any $\gamma\in\{0,1\}^{p}$, let $\gamma_{1:p-1}=(\gamma_1,\gamma_2,\ldots,\gamma_{p-1})\in \{0,1\}^{p-1}$, and let $\bar{\rho}'$ and $\bar{\blam}'$ be mappings defined on $\om_{\M}^{(-p)}$ such that for any $\gamma\in\om_{\M}^{(-p)}$, if $|\gamma_{1:p-1}| < p-1$,
\[
\bar{\rho}'(\gamma) = \rho'(\gamma_{1:p-1}), \quad \bar{\lambda}'_j(\gamma) = \lambda'_j(\gamma_{1:p-1}) \quad \text{for $j=1,2,\ldots,p-1$}, \quad \text{and} \quad  \bar{\lambda}'_p(\gamma)=0,
\]
while if $|\gamma_{1:p-1}|=p-1$, then
\[
\bar{\rho}'(\gamma) = 1 \quad,   \bar{\lambda}'_j(\gamma) = 0 \quad \text{for $j=1,2,\ldots,p-1$}, \quad \text{and} \quad  \bar{\lambda}'_p(\gamma)=1.
\]

Now consider the pFS procedure with $p$ predictors with mappings $\rho$ and $\blam$ defined such that
\begin{itemize}
\item[(i)] If $\gamma \in \om_{\M}^{(-p)}$ and $\pi(\gamma^{+p})>0$,
\begin{align*}
\rho(\gamma) &= \bar{\rho}'(\gamma)\cdot \frac{\pi(\gamma)}{\pi'(\gamma)}\\
\lambda_j(\gamma) &= \left\{ \begin{array}{ll}
\frac{1-\bar{\rho}'(\gamma)}{1-\rho(\gamma)}\cdot \bar{\lambda}'_{j}(\gamma) & \mbox{for $j=1,2,\ldots, p-1$,} \\\\
\frac{\pi(\gamma^{+p})}{\pi'(\gamma)} \cdot \frac{\bar{\rho}'(\gamma)}{1-\rho(\gamma)} & \mbox{for $j=p$,}
\end{array}
\right.
\end{align*}
\item[(ii)] If $\gamma\in \om_{\M}^{(-p)}$ and $\pi(\gamma^{+p})=0$,
\[
\rho(\gamma) = \bar{\rho}'(\gamma) \quad \text{and} \quad \lambda_j(\gamma) = \bar{\lambda}'_j(\gamma) \quad \text{for $j=1,2,\ldots,p$.}
\]
\item[(iii)] If $\gamma \in \om_{\M}\setminus \om_{M}^{(-p)}$ and $|\gamma|<p$,
\[ \rho(\gamma)=1 \quad \text{and} \quad \lambda_j(\gamma)=1/(p-|\gamma|)\cdot \I_{\gamma_j=0}. \]
\end{itemize}
Under this pFS procedure, the $p$th predictor is always the last to be added. Now let us check that the marginal distribution of the final model $\gamma^{(p)}$ is indeed $\pi$.
Now, for any $\gamma\in \om_{\M}^{(-p)}$ such that $\pi(\gamma)>0$, by (i), (ii), and (iii) we have
\begin{align*}
&\sum_{\gamma_{(1)},\ldots,\gamma_{(p)}} \left\{\prod_{t=1}^{|\gamma|}[1-\rho(\gamma_{(t-1)})] \cdot \lambda_{j_t}(\gamma_{(t-1)})\right\}\cdot \rho(\gamma)\\
=&\sum_{\gamma_{(1)},\ldots,\gamma_{(p)}} \left\{\prod_{t=1}^{|\gamma|}[1-\bar{\rho}'(\gamma_{(t-1)})] \cdot \bar{\lambda}'_{j_t}(\gamma_{(t-1)})\right\}\cdot \bar{\rho}'(\gamma)\cdot \pi(\gamma)/\pi'(\gamma)\\
=&\pi'(\gamma)\cdot\pi(\gamma)/\pi'(\gamma)=\pi(\gamma),
\end{align*}
where $j_1,j_2,\ldots,j_{|\gamma|}$ are the values of the selection variables $J_1,J_2,\ldots,J_{|\gamma|}$ that corresponds to the sequence of models $\gamma_{(1)},\ldots,\gamma_{(|\gamma|-1)},\gamma_{(|\gamma|)}=\gamma$. Similarly, for any $\gamma\in\om_{\M}\backslash\om^{(-p)}$ such that $\pi(\gamma)>0$, by (i), (ii), and (iii), the marginal probability for $\gamma_{(p)}$ to be $\gamma$ is
\begin{align*}
&\sum_{\gamma_{(1)},\ldots,\gamma_{(p)}:\gamma_{(p)}=\gamma} \left\{\prod_{t=1}^{|\gamma|}[1-\rho(\gamma_{(t-1)})] \cdot \lambda_{j_t}(\gamma_{(t-1)})\right\}\cdot \rho(\gamma)\\
=&\sum_{\gamma_{(1)},\ldots,\gamma_{(p)}:\gamma_{(p)}=\gamma} \left\{\prod_{t=1}^{|\gamma|-1}[1-\bar{\rho}'(\gamma_{(t-1)})] \cdot \bar{\lambda}'_{j_t}(\gamma_{(t-1)})\right\}\cdot [1- \bar{\rho}(\gamma_{|\gamma|-1})]\cdot \frac{\pi(\gamma)}{\pi'(\gamma_{(|\gamma|-1)})} \cdot \frac{\bar{\rho}'(\gamma_{(|\gamma|-1)})}{[1-\rho( \gamma_{(|\gamma|-1)})]} \cdot 1\\
=&\pi(\gamma)/\pi'(\gamma_{(|\gamma|-1)})\cdot \pi'(\gamma_{(|\gamma|-1)})=\pi(\gamma).
\end{align*}
The second equality follows because for $\gamma\in\om_{\M}\backslash\om^{(-p)}$ such that $\pi(\gamma)>0$, under (i), (ii), and (iii)
\[
\sum_{\gamma_{(1)},\ldots,\gamma_{(p)}:\gamma_{(p)}=\gamma} \left\{\prod_{t=1}^{|\gamma|-1}[1-\bar{\rho}'(\gamma_{(t-1)})] \cdot \bar{\lambda}'_{j_t}(\gamma_{(t-1)})\right\}\cdot \bar{\rho}'(\gamma_{(|\gamma|-1)})=\pi'(\gamma_{(|\gamma|-1)}),
\]
and with probability 1, $\gamma_{(|\gamma|-1)}$ is the model with the $p$th predictor removed from $\gamma$.
\end{proof}

\begin{proof}[Proof of Theorem~3]
Let $S_1,J_1,S_2,J_2,\ldots,S_p,J_p$ be the latent decision variables of the pFS representation of $\pi$ under consideration. We let $(\om_{d},\F_{d})$ be the probability space on which these decision variables are jointly defined. The sequence of models $\gamma_{(1)},\gamma_{(2)},\ldots,\gamma_{(p)}$ are functions of the decision variables and thus also measurable with respect to $(\om_{d},\F_{d})$.

Fixing the data $\D$, the marginal likelihood under the final model, $p(\D\,|\,\gamma_{(p)})$, is also a random variable on $(\om_{d},\F_{d})$.
For any $\gamma \in \om_{\M}$, we define an event $U_{\gamma}$ on $(\om_{d},\F_{d})$ that $\gamma$ is a submodel of the final model $\gamma_{(p)}$, that is, $\gamma_{(p)}$ contains all of the predictors included in $\gamma$. Mathematically, this event can be expressed as
\begin{align*}
U_{\gamma}:&=\{\omega \in \om_{d}: \gamma_{(t)}(\omega)=\gamma \text{ for $t=|\gamma|$}\}.
\end{align*}
Next, we define a mapping $\Phi: \om_{\M}\rightarrow \real$ as follows. For each $\gamma\in\om_{\M}$,
\[
\Phi(\gamma) := E_{\gamma_{(p)}} \left[ \, p(\D\,|\,\gamma_{(p)})\, |\, U_{\gamma} \, \right],
\]
where the data $\D$ is fixed and the expectation is taken over the final model $\gamma_{(p)}$, or equivalently the decision variables, conditional on the event $U_{\gamma}$.

Now for any $\gamma \in\om_{\M}$, we claim that
\[
\Phi(\gamma)=\begin{cases} p(\D\,|\,\gamma) & \text{if $|\gamma|=p$,} \\
  \rho(\gamma)\,p(\D\,|\,\gamma) + (1-\rho(\gamma))\cdot \sum_{j:\gamma_j=0}\lambda_j(\gamma)\,\Phi(\gamma^{+j}), & \text{if $|\gamma|<p$.}
  \end{cases}
\]
To see this, note that if $|\gamma|=p$, then conditional on $U_{\gamma}$, we have $\gamma_{(p)}=\gamma$, and so $E_{\gamma_{(p)}}\,[\,p(\D\,|\,\gamma_{(p)})\,|\,U_{\gamma}\,]=p(\D\,|\,\gamma)$. Now if $|\gamma|=t<p$, then by the tower property,
\begin{align*}
&E_{\gamma_{(p)}}\,[\,p(\D\,|\,\gamma_{(p)})\,|\,U_{\gamma}\,]\\
=&E_{\gamma_{(p)}}\,[\,E_{\gamma_{(p)}}\,[\,p(\D\,|\,\gamma_{(p)})\,|\,S_{t+1},U_{\gamma}\,]\,|\,U_{\gamma}\,]\\
=&E_{\gamma_{(p)}}\,[\,p(\D\,|\,\gamma_{(p)})\,|\,S_{t+1}=1,U_{\gamma}\,]\cdot P(S_{t+1}=1\,|\,U_{\gamma})\\
&+ \sum_{j:\gamma_j=0} E_{\gamma_{(p)}}\,[\,p(\D\,|\,\gamma_{(p)})\,|\,J_{t+1}=j,S_{t+1}=0,U_{\gamma}\,]\cdot P(J_{t+1}=j\,|\,S_{t+1}=0,U_{\gamma})\cdot P(S_{t+1}=0\,|\,U_{\gamma}).
\end{align*}
Now note that $S_{t+1}=1$ and $U_{\gamma}$ together imply that $\gamma_{(p)}=\gamma$ and so
\[
E_{\gamma_{(p)}}\,[\,p(\D\,|\,\gamma_{(p)})\,|\,S_{t+1}=1,U_{\gamma}\,] = p(\D\,|\,\gamma).
\]
Also, for each $j$ such that $\gamma_j=0$, $\{ \omega\in \om_{d}: J_{t+1}(\omega)=1,S_{t+1}(\omega)=0 \} \cap U_{\gamma} \subset U_{\gamma^{+j}}$. Moreover, conditional on the event $U_{\gamma^{+j}}$, $\gamma_{(p)}$ is a function of $S_{t+2},J_{t+2},\ldots,S_{p},J_{p},S_{p+1}$ and so is independent of $S_1,J_1,\ldots,S_{t+1},J_{t+1}$. Thus,
\begin{align*}
E_{\gamma_{(p)}}\,[\,p(\D\,|\,\gamma_{(p)})\,|\,J_{t+1}=j,S_{t+1}=0,U_{\gamma}\,] &= E_{\gamma_{(p)}}\,[\,p(\D\,|\,\gamma_{(p)}) \,|\,J_{t+1}=j,S_{t+1}=0,U_{\gamma},U_{\gamma^{+j}}\,]\\
&= E_{\gamma_{(p)}}\,[\,p(\D\,|\,\gamma_{(p)})\,|\, U_{\gamma^{+j}} \,]\\
&=\Phi(\gamma^{+j}).
\end{align*}
Finally, since $P(S_{t+1}=1\,|\,U_{\gamma})=\rho(\gamma)$ and $P(J_{t+1}=j\,|\,S_{t+1}=0,U_{\gamma})=\lambda_j(\gamma)$, putting the pieces together we have
\[
\Phi(\gamma) = \rho(\gamma)\,p(\D\,|\,\gamma) + (1-\rho(\gamma))\sum_{j:\gamma_j=0}\lambda_j(\gamma)\,\Phi(\gamma^{+j}).
\]
This establishes the above claim about $\Phi$.

Given the mapping $\Phi$, we are now ready to establish the theorem. First, because under the pFS representation the data generative mechanism essentially forms an HMM by Theorem~1, the model space posterior has a pFS representation with the mappings $\rho(\cdot\,|\,\D)$ and $\blam(\cdot\,|\,\D)$ determined by the posterior distributions of the decision variables $S_1,J_1,\ldots,S_p,J_p$. So our proof strategy now is to simply find the posterior distributions of these decision variables.

For any model $\gamma\in \om_{\M}$ with $|\gamma|=t<p$,
\begin{align*}
\rho(\gamma\,|\,\D)= P(S_{t+1}=1\,|\,U_{\gamma},\D) &= \frac{P(S_{t+1}=1,\D\,|\,U_{\gamma})}{P(\D\,|\,U_{\gamma})}\\
&=\frac{E_{\gamma_{(p)}}\,[\,p(\D\,|\,\gamma_{(p)})\,|\,S_{t+1}=1,U_{\gamma}\,]\cdot P(S_{t+1}=1\,|\,U_{\gamma})}{E_{\gamma_{(p)}}\,[\,p(\D\,|\,\gamma_{(p)})\,|\,U_{\gamma}\,]}\\
&=\rho(\gamma) p(\D\,|\,\gamma)/\Phi(\gamma),
\end{align*}
 which is equal to 1 when $\rho(\gamma)=1$. Similarly, if $\rho(\gamma)\neq 1$, then
 \begin{align*}
 \lambda_j(\gamma\,|\,\D) &= P(J_{t+1}=j\,|\,U_{\gamma},S_{t+1}=0,\D)\\
 &=\frac{P(J_{t+1}=j,S_{t+1}=0,\D\,|\,U_{\gamma})}{P(S_{t+1}=0,\D\,|\,U_{\gamma})}\\
 &=\frac{P(J_{t+1}=j,S_{t+1}=0,\D\,|\,U_{\gamma})}{P(\D\,|\,U_{\gamma}) - P(S_{t+1}=1,\D\,|\,U_{\gamma})}\\
 &=\frac{E_{\gamma_{(p)}}\,[\,p(\D\,|\,\gamma_{(p)})\,|\,J_{t+1}=j,S_{t+1}=0,U_{\gamma}\,]\cdot P(J_{t+1}=j\,|\,S_{t+1}=0,U_{\gamma})\cdot P(S_{t+1}=0\,|\,U_{\gamma})}{E_{\gamma_{(p)}}\,[\,p(\D\,|\,\gamma_{(p)})\,|\,U_{\gamma}\,]-E_{\gamma_{(p)}}\,[\,p(\D\,|\,\gamma_{(p)})\,|\,S_{t+1}=1,U_{\gamma}\,]\cdot P(S_{t+1}=1\,|\,U_{\gamma})}\\
 &=\frac{\Phi(\gamma^{+j}) \cdot \lambda_j(\gamma)\cdot (1-\rho(\gamma))}{\Phi(\gamma) - p(\D|\,\gamma) \cdot \rho(\gamma)}.
 \end{align*}
 On the other hand, for $\rho(\gamma)=1$, then given $U_{\gamma}$, $S_{t+1}=0$ with probability 0 and the value of $J_{t}$ for $t > |\gamma|$ has no impact on the final model $\gamma_{(p)}$. So we can simply set
 \[
 \lambda_j(\gamma\,|\,\D) = \lambda_j(\gamma) \quad \text{for all $j$.}
 \]
 The theorem now follows by letting $\phi(\gamma) = \Phi(\gamma)/p(\D\,|\,\bzero)$.
\end{proof}

\section*{S2. Bayes factors under $g$ and hyper-$g$ priors}

For many common priors on the regression coefficients, the BF term in the weight update can be computed either in closed form or well approximated numerically. Here let us consider two popular priors---the $g$-prior and the hyper-$g$ prior.

Given a particular model ${\gamma}$, Zellner's $g$-prior in its most popular form is the following prior on the regression coefficients and the noise variance
\begin{align*}
p(\varphi) \propto 1/\varphi \quad \text{and} \quad \bbet_{\gamma} \,|\, \varphi, {\gamma} \sim N(\bbet_{\gamma}^0,g(\bX^T\bX)^{-1}/\varphi)
\end{align*}
where $\bbet_{\gamma}^0$ and $g$ are hyperparameters. Following the exposition in \cite{liang:2008}, we assume without loss of generality that the predictor variables $\bX_1$, $\bX_2$, \ldots, $\bX_p$ have all been mean centered at zero. Then we can place a common non-informative flat prior on the intercept $\alpha$ for all models. So
$p(\alpha,\varphi) \propto 1/\varphi$. Under this prior setup, one can show that the BF for a model $\gamma$ versus the null model is given by
\[
\BF_0(\gamma)=\frac{(1+g)^{(n-1-|\gamma|)/2}}{\left(1+g(1-R_{\gamma}^2)\right)^{(n-1)/2}}
\]
where $R_{\gamma}^2$ is the coefficient of determination for model $\M_{\gamma}$.

To avoid undesirable features of the $g$-priors such as Barlett's paradox and the information paradox \citep{berger:2001}, \cite{liang:2008} proposed the use of mixtures of $g$-priors. In particular, they introduced the hyper-$g$ prior, which puts the following hyperprior on $g$:
\[
\frac{g}{1+g} \sim {\rm Beta}(1,a/2-1).
\]
This prior also renders a closed form representation for the model-specific marginal likelihood, and thus for the corresponding BFs. In particular, \cite{liang:2008} showed that the BF of a model $\M_{\gamma}$ versus the null model is given by
\[
\BF_0(\gamma)=\frac{a-2}{|\gamma|+a-2}\cdot\ _2F_1\left((n-1)/2, 1; (|\gamma|+a)/2; R_{\gamma}^2\right)
\]
where $_2F_1(\cdot,\cdot;\cdot;\cdot)$ is the hypergeometric function. More specifically, in the notations of \cite{liang:2008},
\[
_2F_1(a,b;c;z)=\frac{\Gamma(c)}{\Gamma(b)\Gamma(c-b)}\int_{0}^{1}\frac{t^{b-1}(1-t)^{c-b-1}}{(1-tz)^{a}}dt.
\]

Therefore, with either the $g$-prior and the hyper-$g$ prior the BF in the weight update can be computed as
\[
\BF\left(\gamma^{i}_{(t)},\gamma^{i}_{(t-1)}\right)=\frac{\BF_0\left(\gamma^{i}_{(t)}\right)}{\BF_0\left(\gamma^{i}_{(t-1)}\right)}.
\]

\section*{S3. Incorporating dilution under model space redundancy}
\label{sec:dilution}
In this subsection we show that the pFS representation allows us much flexibility in incorporating prior information, and we illustrates this through an interesting phenomenon called the dilution effect first noted by \cite{george:1999}. ``Dilution'' occurs when there is redundancy in the model space. More specifically, consider the scenario where there is strong correlation among some of the predictors, and any one of these predictors captures virtually all of the association between them and the response. In this case models that contain different members of this class but are otherwise identical are essentially the same. As a result, if, say, a symmetric prior specification is adopted, these models will receive more prior probability than they properly should. At the same time, other models that do not include members of this class will be down-weighted in the prior. In real data, this phenomenon occurs to varying degrees depending on the underlying correlation structure among the predictors. 

Next, we present a very simple specification of the model space prior under the pFS representation that can effectively address this phenomenon. We do not claim that this approach is the ``best'' way to deal with dilution, but rather use this as an example to illustrate the flexibility rendered by the pFS representation. The specification can most simply be described in two steps. 
\vspace{0.5em}

\noindent {\em Step I. Pre-clustering the predictors based on their correlation.} First, we carry out a hierarchical clustering over the predictor variables using the (absolute) correlation as the similarity metric, which divides the predictors into $K$ clusters---$C_1,C_2,\ldots,C_K$. We recommend using complete linkage for this purpose as this will ensure that the variables within each cluster are all very ``close'' to each other. One need to choose a correlation threshold $s$ for cutting the corresponding dendrogram into clusters---in the case of complete linkage, this is the minimum correlation for two variables to be in the same cluster. We recommend choosing a large $s$, such as 0.9, to place variables into the same basket only if they are very highly correlated.
\vspace{0.5em}

\noindent{\em Step II. Prior specification given the predictor clusters.} Based on the predictor clusters, we assign prior selection probabilities for a model $\M_{\bgam}$ to the variables not yet in the model in the following manner. First, we place equal total prior selection probability over each of the available clusters. Then within each cluster, we assign selection probability evenly across the variables. 
\vspace{0.5em}

For example, consider the situation where there are a total of 10 predictors $X_1$ through $X_{10}$, and following Step~I, they form four clusters $C_1=\{X_1,X_2,X_3\}$, $C_2=\{X_4,X_{10}\}$, $C_3=\{X_5,X_7,X_{9}\}$ and $C_4=\{X_6,X_8\}$. Let $\M_{\bgam}$ be the model that contains variables $X_1$, $X_4$, $X_5$, $X_6$, and $X_8$. That is, $\bgam=(1,0,0,1,1,1,0,1,0,0)$. If the FS procedure reaches $\M_{\bgam}$ and the procedure does not stop, that is, $S(\bgam)=0$, then five variables, $X_2,X_3,X_{7},X_{9},X_{10}$, from three clusters $C_1'=\{X_2,X_3\}$, $C_2'=\{X_4\}$, and $C_3'=\{X_5\}$ are available for further inclusion. In this case we choose the selection probabilities $\blam(\bgam)$ to be: $\lambda_1(\bgam)=\lambda_4(\bgam)=\lambda_5(\bgam)=\lambda_6(\bgam)=\lambda_8(\bgam)=0$, $\lambda_2(\bgam)=\lambda_3(\bgam)=1/3\times 1/2=1/6$, $\lambda_4=1/3$ and $\lambda_5=1/3$. Under such a specification, the predictors falling in the same cluster evenly share a fixed piece of the prior selection probability, which ensures that the prior weight on the other variables are not ``diluted''.

\bibliography{ModelSel}